\documentclass{tlp}
\usepackage{aopmath}
\usepackage{amssymb}
\usepackage{graphics}
\usepackage{color}
\usepackage{comment}
\usepackage{latexsym}
\usepackage{url}
\usepackage{comment}
\usepackage[defblank]{paralist}

\newcommand{\A}{\mathcal{A}} 
\newcommand{\C}{\mathcal{C}} 
 
 \renewcommand{\H}{\mathcal{H}}
 
 \renewcommand{\L}{\mathcal{L}}
 \newcommand{\N}{\mathcal{N}}
 
 \newcommand{\R}{\mathcal{R}}

\newcommand{\ra}{\rightarrow}

\newcommand{\la}{\leftarrow}

\newcommand{\dd}[2]{#1_1,\ldots,#1_{#2}}             

\newcommand{\tup}[1]{\langle #1\rangle}            

\newcommand{\wrt}[0]{with respect to\ }

\renewcommand{\paragraph}[1]{\textbf{#1}}

\newcommand{\adom}[1]{\mathrm{dom}(#1)}

  {$\Box$}

\newcommand{\iso}{\sim}

\newcommand{\rel}[1]{\mathsf{#1}}

\newcommand{\const}[1]{\mathit{#1}}
\newcommand{\vett}[1]{\bar{#1}}

\newcommand{\ext}[2]{#1^{#2}}

\renewcommand{\dom}{\Gamma} 
\newcommand{\freshdom}{\Gamma_f}
\newcommand{\variables}{\Gamma_V}

\newcommand{\eer}[1]{\mathsf{#1}}

\newcommand{\entities}{\mathit{Ent}}
\newcommand{\relationships}{\mathit{Rel}}
\newcommand{\attributes}{\mathit{Att}}
\newcommand{\symbols}{\mathit{Sym}}

\newcommand{\dep}{\Sigma}
\newcommand{\idep}{\Sigma_I}

\newcommand{\kdep}{\Sigma_K}
\newcommand{\rdep}{\Sigma_R}

\newcommand{\key}[1]{\mathit{key}(#1)}
\newcommand{\isa}[1]{\mathit{ISA}}

\newcommand{\head}[1]{\mathit{head}(#1)}
\newcommand{\body}[1]{\mathit{body}(#1)}
\newcommand{\arity}[1]{\mathit{arity}(#1)}

\newcommand{\ans}[3]{\mathit{ans}(#1,#2,#3)} 
\newcommand{\ansf}[3]{\mathit{ans}_f(#1,#2,#3)} 

\newcommand{\Qeq}{q_\mathit{eq}}

\newcommand{\ins}[1]{\bar{#1}}

\newcommand{\insX}{\ins{X}}
\newcommand{\insY}{\ins{Y}}

\newcommand{\varsX}{\bar{X}}

\newcommand{\terms}{\bar{t}}

\newcommand{\sizeConnected}[0]{\lambda_D}
\newcommand{\chase}[2]{\mathit{chase}_{#1}(#2)}
\newcommand{\fchase}[2]{\mathit{chase}^{\maxlevel}_{#1}(#2)}

\newcommand{\level}[1]{\mathit{level}(#1)}

\newcommand{\maxlevel}[0]{\delta_{M}}
\newcommand{\chaseeq}[2]{\mathit{chase^{\predeq}}_{#1}(#2)}
\newcommand{\subtree}[2]{\mathit{subtree}(#1,#2)}

\newcommand{\atom}[1]{\underline{#1}}  

\newcommand{\prog}{\Pi}      
\newcommand{\progid}{\Pi^{\idep}} 
\newcommand{\progdc}{\Pi^{\mathit{DC}}} 
\newcommand{\progkd}{\Pi^{\kdep}} 
\newcommand{\progeq}{\Pi^{\mathit{eq}}}
\newcommand{\progfin}[2]{\Pi^{#1,#2}}

\newcommand{\progbase}{\Pi^{\mathit{ba}}}      

\newcommand{\predeq}[0]{\mathit{eq}}

\newcommand{\pmil}{\la}

\newcommand{\progff}{\prog^{\mbox{\tiny \textit{ff}}}}
\newcommand{\nofresh}[1]{#1^{[\dom]}}

\newcounter{cefalo}

\newcounter{cefalocont}

\newtheorem{definitionAux}{Definition} 
\newenvironment{definition}{\begin{definitionAux} 
}{\hfill\markfull\end{definitionAux}}

\newtheorem{exampleAux}{Example}
\newenvironment{example}{\begin{exampleAux}\upshape}
  {\hfill\markfull\end{exampleAux}}
\newenvironment{exampleCont}[1]{\trivlist
  \item[\hskip \labelsep{\textit{Example~#1 (cont.)}}]\item}{\hfill\markfull\endtrivlist}

\def\qed{\hfill{\qedboxempty}      
  \ifdim\lastskip<\medskipamount \removelastskip\penalty55\medskip\fi}

\def\qedboxempty{\vbox{\hrule\hbox{\vrule\kern3pt
                 \vbox{\kern3pt\kern3pt}\kern3pt\vrule}\hrule}}

\def\qedfull{\hfill{\qedboxfull}   
  \ifdim\lastskip<\medskipamount \removelastskip\penalty55\medskip\fi}

\def\qedboxfull{\vrule height 4pt width 4pt depth 0pt}

\newcommand{\markfull}{\qedboxfull}
\newcommand{\markempty}{\qed}

\newcommand{{\incolumn}}[1]{\begin{tabular}[c]{c} #1 \end{tabular}}
\newcommand{{\incolumnmath}}[1]{\begin{array}[c]{c} #1 \end{array}}

\excludecomment{tobechecked}
\excludecomment{unused}

\begin{document}
\sloppy

\bibliographystyle{acmtrans}

\title[Querying Incomplete Data over Extended ER Schemata]{Querying Incomplete Data\\over Extended ER Schemata}
\author[A. Cal\`\i\, and D. Martinenghi]
{ANDREA CAL\`I\\
Computing Laboratory, University of Oxford\\
Eagle House, Walton Well Road -- Oxford OX2 6ED, United Kingdom\\
\email{andrea.cali@comlab.ox.ac.uk}
\and DAVIDE MARTINENGHI\\
Dipartimento di Elettronica e Informazione, Politecnico di Milano\\
Piazza Leonardo 32 -- 20133 Milano, Italy\\
\email{davide.martinenghi@polimi.it}
}

\pagerange{\pageref{firstpage}--\pageref{lastpage}}
\volume{\textbf{??} (??):}
\jdate{??? ???}
\setcounter{page}{1}
\pubyear{????}

\maketitle

\label{firstpage}

\begin{abstract} 
Since Chen's Entity-Relationship (ER) model, conceptual modeling has been playing a fundamental role in relational data design. In this paper we consider an extended ER (EER) model enriched with cardinality constraints, disjointness assertions, and is-a relations among both entities and relationships.
In this setting, we consider the case of incomplete data, which is likely to occur, for instance, when data from different sources are integrated. In such a context, we address the problem of providing correct answers to conjunctive queries by reasoning on the schema. Based on previous results about decidability of the problem, we provide a query answering algorithm that performs rewriting of the initial query into a recursive Datalog query encoding the information about the schema. We finally show extensions to more general settings. This paper will appear in the special issue of {\emph Theory and Practice of Logic Programming (TPLP)} titled  \emph{Logic Programming in Databases: From Datalog to Semantic-Web Rules}.
\end{abstract}

\begin{keywords}
	Extended ER model, Dependencies, Chase, Incomplete Data
\end{keywords}

\section{Introduction}
\label{sec:introduction}

Conceptual data models, and in particular the Entity-Relationship (ER) model~\cite{Chen76}, have long been playing a fundamental role in database design.  With the emerging trends in data exchange, information integration, semantic web, and web information systems, the need for dealing with inconsistent and incomplete data has arisen.
In this context, it is important to provide correct answers to queries posed over inconsistent and incomplete data~\cite{ArBC99}.  It is worth noticing here that inconsistency and incompleteness of data is considered with respect to a set of constraints (a.k.a.{} data dependencies). Such constraints, rather than expressing properties that hold on the data, are used to represent properties of the domain of interest.

We address the problem of answering queries over \emph{incomplete data}, where queries are conjunctive queries expressed over particular relational schemata, called \emph{conceptual schemata}, that are derived from conceptual models.
As for the conceptual models, we follow~\cite{Chen76}, and we adopt an extension of the well-known Entity-Relationship model, that we call \emph{Extended Entity-Relationship (EER) Model}, along with~\cite{Thalheim:2000} and the many variants of the classical ER Model.
Such an extension is widely adopted in practice and is able to represent classes of objects with their attributes, relationships among classes, cardinality constraints in the participation of entities in relationships, and is-a relations among both classes and relationships.
We provide a formal semantics to our conceptual model in terms of the relational database model, similarly to what is done in~\cite{MaMa90}.
This allows us to formulate conjunctive queries over EER schemata.
We do this by providing a translation from EER into relational, whose purpose is to obtain a precise characterization of the relational dependencies that are derived from an EER schema in a design process.

In the presence of data that are incomplete w.r.t.{} to a set of constraints, we need to \emph{reason} about the dependencies in order to provide certain answers; we do this in a model-theoretic fashion, following the approach of~\cite{ArBC99,CCDL01e}.
Intuitively, we start from a given, incomplete database for the relational schema associated with the EER schema; such data, together with the constraints, are interpreted as a logical theory, with a (possibly infinite) set of models, also called \emph{solutions} in the literature.
We adopt the so-called \emph{sound semantics} (see, e.g., \cite{CaLR03}): a database is a model if it is a superset of the initial data, and satisfies the constraints.
Given a query, the \emph{certain answers} are those that are true in all models.

In this paper we address the problem of answering conjunctive queries over schemata derived from EER schemata in the presence of incomplete data with respect to the schema under the sound semantics.
We present an algorithm, based on encoding the information about the conceptual schema and the instance into a \emph{rewriting} of the conjunctive query in Datalog, which computes the certain answers to queries posed in such a context.
The algorithm reasons on the integrity constraints and the query.

The problem at hand can be sketchily stated as follows.
\begin{compactitem}
	\item We have a conceptual EER schema. From it, a relational schema $S$ is obtained through a translation mechanism that also produces a set of integrity constraints $\Sigma$ consisting of key and inclusion dependencies.
	\item We also have an instance $D$ for $S$. $D$ may be inconsistent with respect to $\Sigma$ and incomplete.
	\item Consider all the $S$-instances that extend $D$ and satisfy $\Sigma$.
	The certain answers to a conjunctive query $Q$ over $S$ are those that are true of all those instances.
	\item The problem is how to compute the certain answers to $Q$.
	\item The solution we propose is to translate $Q$ into a new query $Q^*$ and pose it to $D$.
	The answers to $Q^*$ are the certain answers to $Q$.
\end{compactitem}

More specifically, our contribution is summarized as follows.
\begin{compactenum}[\itshape (a)]
	\item We define a class of relational dependencies, that we call \emph{conceptual dependencies (CDs)} that is able to represent EER schemata; our class is constituted by a subset of the well-known \emph{key dependencies (KDs)} and \emph{inclusion dependencies (IDs)}.
	A broad class of KDs and IDs for which the query answering problem under incomplete data is known to be decidable is the class of KDs (at most one per relational predicate) and \emph{non-key-conflicting inclusion dependences (NKCIDs)}, that was introduced in~\cite{CaLR03}.
	The problem of answering incomplete data under general KDs and IDs is known to be undecidable~\cite{CaLR03}.
	\item We tackle the problem of query answering under CDs in the presence of incomplete information, under the sound semantics.
	After reviewing how, also under CDs, the chase is a useful tool for query answering, we solve the problem by means of query rewriting, in the same fashion as in~\cite{CaLR03b}, where a rewriting for KDs and NKCIDs is presented.
	We show an algorithm that, given a query, rewrites it into another one that encodes relevant information about the relational constraints, so that the evaluation of the rewritten query over the initial incomplete data returns the certain answers.
	The rewritten query is in (positive) Datalog.
\end{compactenum}
                                                  
Note that the chase (which we, however, do not construct in our query answering technique) is a conceptual tool whose construction amounts to repairing violations of IDs and KDs, the former by adding tuples, and the latter by merging tuples. However, repairing is not always possible, and in such cases the chase does not exist and query answering becomes trivial. In such cases the repair would require tuple deletions: this is captured by semantics such as those in \cite{DBLP:conf/dagstuhl/BertossiB05,CaLR03b}.

It is important to notice that the class of CDs does not fall into the class of KDs and NKCIDs.
A strong indication (though there is no formal proof) of the decidability, that we show in this paper, of the query answering problem under CDs (and under the sound semantics) is found in~\cite{CaDL98}, where it is shown that query containment in a description logic, capable of representing EER schemata, is decidable.
However, the technique of~\cite{CaDL98} does not give any indication on the algorithm that may be used to check containment (or, in our case, to answer queries).  Differently, our technique gives a direct tool for query answering that, under certain conditions on the data, provides a low computational complexity with respect to the size of the data.

This paper extends the work in~\cite{DBLP:conf/er/Cali07} and is organized as follows.
We give necessary preliminaries in Section~\ref{sec:preliminaries}; we introduce the EER model in Section~\ref{sec:conceptual-model}; in Section~\ref{sec:chase} we show how to answer queries with the \emph{chase}, a formal tool to deal with dependencies; the query rewriting technique is described in~\ref{sec:answering}, together with extensions to more general cases.
Section~\ref{sec:discussion} concludes the paper, discussing related works.

\section{Preliminaries and notation}
\label{sec:preliminaries}

In this section we give a formal definition of the relational data model, database constraints, conjunctive queries and answers to queries on incomplete data.

In the relational data model~\cite{Codd70}, predicate symbols are used to denote the relations in the database, whereas constant symbols denote the objects and the values stored in relations.
We assume to have two distinct, fixed and infinite alphabets $\freshdom$ and $\dom$ of \emph{fresh constants} and \emph{non-fresh constants} respectively, and we consider only databases over $\dom \cup \freshdom$.
We note that
fresh constants are introduced as a technical construct that allows us to build some
representatives of
databases, as will be explained when introducing the chase.
In particular, fresh constants are similar to labeled nulls~\cite{FKMP05} in that they allow representing existentially quantified variables and will thus later be associated with Skolem terms.
Indeed, fresh constants play a role analogous to that of Skolem terms.
For non-fresh constants, which represent the proper constants of the universe, we adopt the so-called \emph{unique name assumption} i.e., we assume that different non-fresh constants denote different objects.
Instead, fresh constants can be thought of as placeholders for non-fresh constants. Therefore, distinct fresh constants can also represent the same object.
Furthermore, we shall make use of variables from a set $\variables$.

A \emph{relational schema} $\R$ consists of an alphabet of \emph{predicate} (or \emph{relation}) symbols, each with an associated \emph{arity} denoting the number of arguments of the predicate (or attributes of the relation).
When a relation symbol $r$ has arity $n$, it can be denoted by $r/n$; in general, the arity of $r$ can also be indicated by $\arity{r}$.

A \emph{relational database} (or simply database) $D$ over a schema $\R$ is a set of relations with constants as atomic values.
We have one relation of arity $n$ for each predicate symbol of arity $n$ in the alphabet $\R$.
The relation $\ext{r}{D}$ in $D$ corresponding to the predicate symbol $r$ consists of a set of tuples of constants, that are the tuples satisfying the predicate $r$ in $D$.

When, given a database $D$ for a schema $\R$, a tuple $t=(\dd{c}{n})$ is in $\ext{r}{D}$, where $r\in\R$, we say that the fact $r(\dd{c}{n})$ holds in $D$.
Henceforth, we will interchangeably use the notion of fact and tuple.

\paragraph{Integrity constraints.}
\emph{Integrity constraints} are assertions on the symbols of the alphabet $\R$ that are intended to be satisfied in every database for the schema.
The notion of satisfaction depends on the type of constraints defined over the schema.

The database constraints of interest are \emph{inclusion dependencies (IDs)} and \emph{key dependencies (KDs)}~(see e.g.~\cite{AbHV95}).
We denote with over-lined uppercase letters (e.g., $\insX$) both sequences and sets of attributes of relations, and enclose them between vertical bars to denote the number of attributes in the set or sequence (e.g., $|\insX|$).
Given a tuple $t$ in relation $\ext{r}{D}$, i.e., a fact $r(t)$ in a database $D$ for a schema $\R$, and a sequence of attributes $\insX$ of $r$, we denote with $t[\insX]$ the \emph{projection}~(see e.g.~\cite{AbHV95}) of $t$ on the attributes in $\insX$.
\begin{compactenum}[\itshape (i)]
	\item \textit{Inclusion dependencies (IDs).}
	An inclusion dependency $\sigma_I$ between relational predicates $r_1$ and $r_2$ is denoted by $r_1[\ins{X}] \subseteq r_2[\ins{Y}]$.
	Given a database $D$ with values only in $\dom$, such a constraint is satisfied in $D$, written $D\models\sigma_I$, iff, for each tuple $t_1$ in $\ext{r_1}{D}$, there exists a tuple $t_2$ in $\ext{r_2}{D}$ such that $t_1[\ins{X}]=t_2[\ins{Y}]$. An ID is said to be a \emph{full-width} ID if every attribute of $r_1$ occurs in $\ins{X}$ exactly once and every attribute of $r_2$ occurs in $\ins{Y}$ exactly once.
	\item \textit{Key dependencies (KDs).}
	A key dependency $\sigma_K$ over a relational predicate $r$ with $\arity{r}\geq 2$ is denoted by $\key{r}=\ins{K}$, where $\ins{K}$ is a nonempty subset of the attributes of $r$.
	Given a database $D$ with values only in $\dom$, such a constraint is satisfied in $D$, written $D\models\sigma_K$, iff, for each $t_1,t_2\in\ext{r}{D}$ such that $t_1\neq t_2$, we have $t_1[\ins{K}^*]\neq t_2[\ins{K}^*]$, where $\ins{K}^*$ is any sequence of $|\ins{K}|$ attributes where each attribute in $\ins{K}$ occurs exactly once.
	Observe that KDs are a special case of functional dependencies (FDs)~\cite{AbHV95}.
	Note also that we restricted our definition to predicates with arity at least $2$, since for predicates of smaller arity keys would be always satisfied (under set semantics).
\end{compactenum}
Above, we specified when dependencies are satisfied in databases with values only in $\dom$. For databases with values in $\dom \cup \freshdom$, we define satisfaction of dependencies as follows. Given a (key or inclusion) dependency $\sigma$ and a database $D$ with values in $\dom \cup \freshdom$, let $B$ be a database obtained from $D$ by replacing every distinct fresh constant with a distinct non-fresh constant that does not appear elsewhere in $D$. We have that $\sigma$ is satisfied in $D$, written $D\models \sigma$, iff $B\models \sigma$.

A database $D$ over a schema $\R$ is said to \emph{satisfy} a set of integrity constraints $\dep$ expressed over $\R$, written $D \models \dep$, if every constraint in $\dep$ is satisfied by $D$.


We now briefly introduce the basics of logic programming and Datalog and refer to \cite{Lloy87} for further details.

\paragraph{Logic programs.}
%
Logic programs are formulated in a language $\L$ of predicates and functions of nonnegative arity; $0$-ary functions are constants.
A language $\L$ is function-free if it contains no functions of arity greater than $0$.
A \emph{term} is inductively defined as follows: each variable $X$ and each constant $c$ is a term, and if $f$ is an $n$-ary function symbol and $t_1, \dots, t_n$ are terms, then $f(t_1, \dots, t_n)$ is a term.
A term is \emph{ground} if no variable occurs in it.
The \emph{Herbrand universe} of $\L$, denoted $U_{\L}$, is the set of all ground terms that can be formed with the functions and constants in $\L$.
An \emph{atom} is a formula $p(t_1, \dots, t_n)$, where $p$ is a predicate symbol of arity $n$ and each $t_i$ is a term; the atom is ground if all $t_i$ are ground.
The \emph{Herbrand base} of a language $\L$, denoted $B_{\L}$, is the set of all ground atoms that can be formed with predicates from $\L$ and terms from $U_{\L}$.
A \emph{definite clause} is a rule of the form
\[
\atom{A}_0 \la \atom{A}_1, \dots, \atom{A}_m\quad (m\geq 0)
\]
where each $\atom{A}_i$ is an atom.
The parts on the left and on the right of ``$\la$'' are called the \emph{head} and the \emph{body} of the rule, respectively.
For a rule $\rho$, we also denote its head by $\head{\rho}$, and its body by $\body{\rho}$.
A rule whose body is empty ($m=0$) and whose head is ground is called a \emph{fact}.
A \emph{logic program} is a set of definite clauses.
A clause or logic program is ground if it contains no variables.
A clause is \emph{range-restricted} if every variable in it also occurs in its body.
A program is range-restricted if all its clauses are.

Each logic program $\Pi$ is associated with the language $\L(\Pi)$ consisting of the predicates, functions, and constants occurring in $\Pi$.
If no constant occurs in $\Pi$, we add some constant to $\L(\Pi)$ to have a nonempty domain.
We simply write $U_{\Pi}$ and $B_{\Pi}$ for $U_{\L(\Pi)}$ and $B_{\L(\Pi)}$, respectively.
A \emph{Herbrand interpretation} of a logic program $\Pi$ is any subset $I \subseteq B_{\Pi}$ of its Herbrand base.
Intuitively, the atoms in $I$ are true, and all others are false.
A \emph{Herbrand model} of $\Pi$ is a Herbrand interpretation of $\Pi$ such that for each rule $\atom{A}_0 \la \atom{A}_1, \dots, \atom{A}_m$ in $\Pi$, this interpretation satisfies the formula $\forall X_1\dots\forall X_n(\atom{A}_1\land \dots\land \atom{A}_m) \ra \atom{A}_0$, where $X_1, \dots, X_n$ are all the variables in the rule.

Let $\Pi$ be a logic program; the \emph{immediate consequence operator} $T_\Pi$ on $\Pi$ is a function from the set of all Herbrand interpretations of $\Pi$ into itself, defined as
\[
T_\Pi(I) = \{ \atom{A}_0 \in B_\Pi \mid \mbox{there is } (\atom{A}_0 \la \atom{A}_1, \dots, \atom{A}_m) \mbox{ in } \Pi \mbox{ and } \{ \atom{A}_1, \dots, \atom{A}_m\} \subseteq I\}
\]
The sequence $T_\Pi^0=\emptyset, T_\Pi^{i+1}=T_\Pi(T_\Pi^i), i\geq 0$ always admits a limit, denoted by $T_\Pi^\infty$, which coincides with the \emph{least Herbrand model} of $\Pi$, i.e., the unique minimal model of $\Pi$ (a model being minimal if no proper subset thereof is also a model).
For a set of (ground or non-ground) clauses $\Pi$, the immediate consequence operator is defined as $T_\Pi=T_{gr(\Pi)}$, where $gr(\Pi)$ is the set of all clauses obtained from any clause in $\Pi$ by substituting elements of $U_\Pi$ for the variables.
A ground atom $\atom{A}$ is called a \emph{consequence} of a set $\Pi$ of clauses if $\atom{A} \in T_\Pi^\infty$, and we write $\Pi \models \atom{A}$.

An $n$-ary query $\Pi_q$ over a schema $\R$ consists of an $n$-ary predicate $q$ (called query predicate) and a finite set $\Pi$ of definite clauses such that
\begin{compactenum}[\itshape (1)]
	\item $q$ is the head predicate for at least one rule in $\Pi$;
	\item the predicate symbols of the head atoms are not relation symbols in $\R$;
	\item the predicate symbols of the body atoms are either relation symbols in $\R$ or one of the head predicates of a rule in $\Pi$.
\end{compactenum}
The evaluation, called \emph{answer}, of a query $\Pi_q$ over a database $D$ (which is a set of facts), written $\Pi_q(D)$, is the restriction to $q$ over the least Herbrand model $M$ of the logic program $\Pi \cup D$, i.e., the largest subset of $M$ containing only atoms with predicate $q$.
It will be made clear by the context whether by $\Pi_q(D)$ we refer to the set of facts or to the set of tuples in the answer.

A \emph{Datalog} clause is a range-restricted definite clause whose terms are either variables or constants
(no function symbols).
A Datalog program is a  set of Datalog clauses.
The notion of query given above also applies to Datalog, since Datalog programs are a specialization of logic programs.

\paragraph{Conjunctive queries.}
In general, a \emph{relational query} is a formula that specifies a set of data to be retrieved from a database.
In the following we will refer to the class of conjunctive queries.
A \emph{conjunctive query} (CQ) of arity $n$ over a schema $\R$ is a Datalog query $\Pi_q$ such that $\Pi$ consists of a single rule in which
\begin{compactenum}[\itshape (1)]
	\item the head is of the form $q(\vett{X})$, where $\vett{X}$ is a sequence of distinct variables;
	\item the constants occurring in the body are from $\dom$;
	\item the predicate symbols of the atoms in the body are in $\R$ ($q$ does not occur in the body).
\end{compactenum}
The variables occurring in the head of a conjunctive query are called \emph{distinguished variables}, the others variables occurring in the body are the \emph{non-distinguished variables}.
For simplicity, the answer to a conjunctive query $q$ over a database $D$ for $\R$ is more compactly denoted as $q(D)$ (rather than $\Pi_q(D)$).

The answers we are mainly interested in are those that contain no fresh constants, because fresh constants merely represent existentially qualintied variables, in the same way as Skolem terms and labeled nulls~\cite{FKMP05}.
Therefore we introduce the notation $\nofresh{q}(D)$ for a CQ $q$ to indicate the largest subset of $q(D)$ whose tuples contain no fresh constants.

\paragraph{Homomorphism.}  
A \emph{mapping} from one set of symbols, $S_1$, to another set of symbols, $S_2$, is a function $\mu: S_1 \ra S_2$ defined as follows:
\begin{inparaenum}[\itshape (i)]
	\item $\emptyset$ (empty mapping) is a mapping;
	\item if $\mu_0$ is a mapping, then $\mu_0 \cup \{X \ra Y\}$, where $X \in S_1$ and $Y \in S_2$ is a mapping if $\mu_0$ does not already contain some $X \ra Y'$ with $Y \neq Y'$.
\end{inparaenum}
If $X \ra Y$ is in a mapping $\mu$, we write $\mu(X) = Y$.
A \emph{homomorphism} from a set of atoms $D_1$ to another set of atoms $D_2$, both over the same relational schema $\R$, is a mapping $\mu$ from $\dom \cup \freshdom \cup \variables$ to $\dom \cup \freshdom \cup \variables$ such that the following conditions hold:
\begin{inparaenum}[\itshape (1)]
	\item if $c \in \dom$ then $\mu(c)=c$; 
	\item if $c \in \freshdom$ then $\mu(c) \in \dom\cup\freshdom$; 
	\item if the atom $r(\dd{c}{n})$ is in $D_1$, then the atom $r(\mu(c_1), \ldots, \mu(c_n))$ is in $D_2$.
\end{inparaenum}
In the following, sometimes a homomorphism may have a codomain different from $\dom \cup \freshdom \cup \variables$; for instance, it could contain terms from the Herbrand universe of a logic program: in such cases, this will be made explicit.

The notion of homomorphism is naturally extended to atoms as follows.
If $\atom{F} = r(\dd{c}{n})$ is an atom and $\mu$ a homomorphism, we define $\mu(\atom{F}) = r(\mu(c_1), \ldots, \mu(c_n))$.
For a \emph{set} of atoms, $F = \{\dd{\atom{F}}{m}\}$, we define $\mu(F) = \{ \mu(\atom{F}_1), \ldots, \mu(\atom{F}_m) \}$.
The set of atoms $\{\mu(\atom{F_1}, \ldots, \mu(\atom{F}_m)\}$ is also called \emph{image} of $F$ \wrt $\mu$.
In this case, we say that $\mu$ \emph{maps} $F$ to $\mu(F)$.
For a \emph{conjunction} of atoms $\Phi = \dd{\atom{F}}{n}$, we use $\mu(\Phi)$ to denote the set of atoms $\mu(\{ \dd{\atom{F}}{n} \})$.
An \emph{isomorphism} is a bijective homomorphism.


\paragraph{Querying incomplete data.}
In the presence of incomplete data, a natural way of considering the problem of query answering is to adopt the so-called \emph{sound semantics} or \emph{open-world assumption}~\cite{Reit78,lenz02}.
In this approach, the data are considered sound but not complete, in the sense that they constitute a piece of correct information, but not necessarily all the relevant information.
In such a case, we need to reason in the presence of incomplete information, thus considering a theory (given by the schema and constraints) having multiple models.
In our context, under relational constraints, it often happens that the data do not satisfy the constraints, especially in information integration, where heterogeneous data are represented by a single schema.
Reasoning with incomplete information allows us to address those constraint violations that are caused by the absence of elements from the database (such as inclusion dependencies). (Note that violations of other kinds of constraints, such as key dependencies, cannot be addressed in this way.)
More formally, we restrict our attention to the so-called \emph{certain answers} to a query: given a finite database $D$, the answers we consider are those that are true in all models, i.e., in \emph{all} the databases that contain $D$ \emph{and} satisfy the dependencies.
In the following, we shall always assume that the initial database has finite size, while no finiteness assumptions is made on the models.
\begin{definition}[Certain answer]\label{def:answers}
	Consider a relational schema $\R$ with a set of dependencies $\dep$, and a finite database $D$ for $\R$.
	Let $q$ be a conjunctive query of arity $n$ over $\R$.
	A $n$-tuple $t$ is a \emph{certain answer} to $q$ w.r.t.~$D$ and $\dep$ if and only if, for every database $B$ for $\R$ such that $B \models \dep$ and $B \supseteq D$, we have $t \in q(B)$, and $t$ consists of constants in $\dom$.
	The set of certain answers is denoted by $\ans{q}{\dep}{D}$.
\end{definition}
\begin{example}\label{exa:certain-answer}
	Consider a relational schema $\R$, here inspired by \cite{CaLR03b}, with the relations $\rel{player}/2$ (player-team pairs) and $\rel{team}/2$ (team-city pairs), a set of IDs $\Sigma = \{ \mathsf{player}[2] ~\subseteq~ \mathsf{team}[1]\}$, and a database $D$ consisting of the facts
	$\rel{player}(\mathit{pirlo},\mathit{acMilan})$,
	$\rel{player}(\mathit{totti},\mathit{roma})$,
	$\rel{team}(\mathit{acMilan},\mathit{milan})$.

	The ID in $\Sigma$ tells us that $\mathit{roma}$ is the name of some team in every database $B\supseteq D$ such that $B\models\Sigma$, i.e., each such database $B$ must contain at least a fact of the form $\rel{team}(\mathit{roma},c)$, where $c$ is some value in $\dom$.

	Consider now the query $q(X) ~\la~ \mathsf{team}(X,Y)$, asking the names of the teams in the database.
	By the above considerations, the set of certain answers is $\{\mathit{acMilan},\mathit{roma}\}$.

	Let $\atom{F}$ be the fact $\rel{team}(\mathit{roma},\alpha)$, where $\alpha$ is a value in $\freshdom$.
	As we will show in Section~\ref{sec:chase}, there is a homomorphism from $D\cup\{\atom{F}\}$ to every database $B'\supset D$ such that $B'\models\Sigma$.
	Consider, e.g., such a database
	$B'=\{ \rel{player}(\mathit{pirlo},\mathit{acMilan}),$
	$\rel{player}(\mathit{totti},\mathit{roma}),$
	$\rel{team}(\mathit{acMilan},\mathit{milan}),$
	$\rel{team}(\mathit{roma},\mathit{rome}),$
	$\rel{team}(\mathit{psg},\mathit{paris})\}$.
	There is a homomorphism $\lambda$ from $D\cup\{\atom{F}\}$ to $B'$ such that
	\begin{inparaenum}[\itshape (i)]
		\item $\lambda(\alpha)=\mathit{rome}$, 
		\item $\lambda(\atom{F})=\rel{team}(\mathit{roma},\mathit{rome})$, 
		\item $\lambda$ sends all facts in $D$ into themselves, and 
		\item $B' = \lambda(D\cup\{\atom{F}\}) \cup \{\rel{team}(\mathit{psg},\mathit{paris})\}$.
	\end{inparaenum}
\end{example}

We will see that, under the database dependencies we consider in this paper, the problem of query answering is mainly complicated by two facts:
\begin{inparaenum}[\itshape (i)]
	\item the number of databases that satisfy $\dep$ and that include $D$ can be infinite; 
	\item there is no bound to the size of such databases.
\end{inparaenum}


\begin{definition}[Querying incomplete databases]\label{def:problem}
	Consider a relational schema $\R$ with a set of dependencies $\dep$, and a finite database $D$ for $\R$.
	Let $q$ be a conjunctive query of arity $n$ over $\R$.
	The problem of querying incomplete databases under $\dep$ is the problem of determining \emph{all} tuples in $\ans{q}{\dep}{D}$.
	The corresponding decision problem is determining, given also a tuple $t$ of arity $n$, whether $t\in\ans{q}{\dep}{D}$.
\end{definition}


\section{The Conceptual Model}
\label{sec:conceptual-model}

In this section we present the conceptual model we shall deal with in the rest of the paper, and we give its semantics in terms of relational database schemata with constraints.

Such model incorporates the basic features of the ER model~\cite{Chen76} and OO~models, including subset (or is-a) constraints on both entities and relationships.
It is an extension of the one presented in~\cite{CCDL01e}, and here we use a notation analogous to that of~\cite{CCDL01e}.
Henceforth, we will call such a model \emph{Extended Entity-Relationship (EER) model}, and we will call schemata expressed in the EER model \emph{Extended Entity-Relationship (EER) schemata}.

An \emph{EER schema} consists of a collection of entity, relationship, and attribute definitions over an \emph{alphabet $\symbols$ of symbols}.
The alphabet $\symbols$ is partitioned into a set of entity symbols (denoted by $\entities$), a set of relationship symbols (denoted by $\relationships$), and a set of attribute symbols (denoted by $\attributes$).

An \emph{entity definition} has the form
\begin{tabbing}
  \hspace*{1cm}\=\quad\=\+\kill
  \textsf{entity} $E$\\
  \>\textsf{isa}: $E_1, \ldots, E_h$\\
  \>\textsf{participates}($\geq 1$): $R_1:c_1, \ldots, 
  R_{\ell}:c_{\ell}$\\
  \>\textsf{participates}($\leq 1$): $R'_1:c'_1, \ldots, 
  R'_{\ell'}:c'_{\ell'}$\\
\end{tabbing}
where:
\begin{inparaenum}[\itshape (i)]
	\item $E \in \entities$ is the entity to be defined; 
	\item the \textsf{isa} clause specifies a set of entities to which $E$ is related via is-a (i.e., the set of entities that are supersets of $e$); 
	\item the \textsf{participates}($\geq 1)$ clause specifies those relationships in which an instance of $E$ must necessarily participate; and for each relationship $R_i$, the clause specifies that $E$ participates as $c_i$-th component in $R_i$;
	\item the \textsf{participates}$(\leq 1)$ clause specifies those relationships in which an instance of $E$ cannot participate more than once (components are specified as in the previous case).
\end{inparaenum}
The \textsf{isa}, \textsf{participates}($\geq 1$) and \textsf{participates}($\leq$ 1) clauses are optional.
Every relationship mentioned in the \textsf{participates}$(\leq 1)$ and \textsf{participates}($\geq 1$) clauses must then be defined accordingly, by mentioning the participating entity as one of the entities of the relationship in a relationship definition.
A \emph{relationship definition} has the form
\begin{tabbing}
  \hspace*{1cm}\=\quad\=\+\kill
  \textsf{relationship} $R$ \textsf{among} $E_1,\ldots, E_n$\\
  \>\textsf{isa}: $R_1[j_{1\,1},\ldots,j_{1\,n}], \ldots, 
  R_h[j_{h\,1},\ldots,j_{h\,n}]$\\
\end{tabbing}
where:
\begin{inparaenum}[\itshape (i)] 
	\item $R\in\relationships$ is the relationship to be defined; 
	\item the $n$ entities of $\entities$, with $n\geq 2$, listed in the \textsf{among} clause are those among which the relationship is defined (i.e., component $i$ of $R$ is an instance of entity $E_i$);
	\item the \textsf{isa} clause specifies a set of relationships to which $R$ is related via is-a; for each relation $R_i$, we specify in square brackets how the components $[1,\ldots,n]$ are related to those of $e_i$, by specifying a permutation $[j_{i\,1}, \ldots, j_{i\,n}]$ of the components of $E_i$; 
	\item the number $n$ of entities in the \textsf{among} clause is the \emph{arity} of $R$.
\end{inparaenum}
The \textsf{isa}, clause is optional.
An \emph{attribute definition} has the form
\begin{tabbing}
  \hspace*{1cm}\=\quad\=\+\kill
  \textsf{attribute} $A$ \textsf{of} $X$ \\
  \>\textsl{qualification}\\
\end{tabbing}
where:
\begin{inparaenum}[\itshape (i)]
	\item $A \in\attributes$ is the attribute to be defined; 
	\item $X$ is the entity or relationship with which the attribute is associated; 
	\item \textsl{qualification} consists of none, one, or both of the keywords \textsf{functional} and \textsf{mandatory}, specifying respectively that each instance of $X$ has a unique value for attribute $A$, and that each instance of $X$ needs to have at least a value for attribute $A$.
\end{inparaenum}
If the \textsf{functional} or \textsf{mandatory} keywords are missing, the attribute is assumed by default to be multivalued and optional, respectively.

For the sake of simplicity, and without any loss of generality, we assume that in our EER~model attributes of entities or relationships have unique names in a schema.
We also assume that every attribute or entity takes values from an infinite domain.

\smallskip


The semantics of an EER schema $\C$ is defined by
\begin{inparaenum}[\itshape (i)]
	\item associating a relational schema $\R$ to it, and 
	\item specifying when a database for $\R$ satisfies all constraints imposed by the constructs of the schema $\C$.
\end{inparaenum}

We now formally define the relational schema associated with an EER diagram.
Such a relational schema is defined in terms of \emph{predicates}, which represent the so-called concepts (entities, relationships, and attributes) of the EER schema.
%
\begin{compactenum}[\itshape (a)]
	\item Each entity $E$ in $\C$ has an associated predicate $e$ of arity~1.
	Informally, a fact of the form $e(c)$ asserts that $c$ is an instance of entity $E$.
	\item Each attribute $A$ for an entity $E$ in $\C$ has an associated predicate $a$ of arity~2.
	Informally, a fact of the form $a(c,d)$ asserts that $d$ is the value of attribute $A$ associated with $c$, where $c$ is an instance of entity $E$.
	\item Each relationship $R$ involving the entities $\dd{E}{n}$ in $\C$ has an associated predicate $r$ of arity $n$.
	Informally, a fact of the form $r(\dd{c}{n})$ asserts that $(\dd{c}{n})$ is an instance of relationship $R$, where $\dd{c}{n}$ are instances of $\dd{E}{n}$ respectively.
	\item Each attribute $A$ for a relationship $R$ among the entities $\dd{E}{n}$ in $\C$ has an associated predicate $a$ of arity $n+1$.
	Informally, a fact of the form $a(c_1,\ldots,c_n,d)$ asserts that $d$ is a value of attribute $A$ associated with the instance $(c_1,\ldots,c_n)$ of relationship $R$.
\end{compactenum}

Notice that, in our particular relational representation, entities are represented
by unary predicates, which can be thus seen as ``surrogate keys'', i.e.,
attributes that are identifiers and do not have any real-world meaning.
With this representation, user-defined key attributes are not necessary.

In the following, the expression ``query over an EER schema $\C$'' will indicate a query over the relational schema associated wih $\C$ according to the above points (a) to (d).

\begin{figure}[tb]
  \centering \input{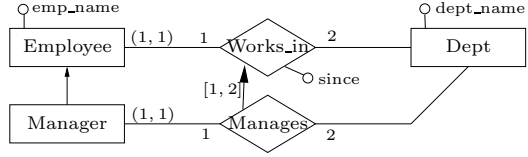}
  \caption{EER schema for Example~\ref{exa:ER-query}}
  \label{fig:ER}
\end{figure}

\begin{example} \label{exa:ER-query}
	Consider the EER schema $\C$ defined as follows.
	\begin{tabbing}
		\hspace*{1cm}\=\quad\=\+\kill
		\textsf{entity} $\rel{Employee}$\\
		\>\textsf{participates}($\geq 1$): $\rel{Works\_in}:1$\\
		\>\textsf{participates}($\leq 1$): $\rel{Works\_in}:1$\\
		\textsf{entity} $\rel{Manager}$\\
		\>\textsf{isa}: $\rel{Employee}$\\
		\>\textsf{participates}($\geq 1$): $\rel{Manages}:1$\\
		\>\textsf{participates}($\leq 1$): $\rel{Manages}:1$\\
		\textsf{entity} $\rel{Dept}$\\
		\textsf{relationship} $\rel{Works\_in}$ \textsf{among} $\rel{Employee},\rel{Dept}$\\
		\textsf{relationship} $\rel{Manages}$ \textsf{among} $\rel{Manager},\rel{Dept}$\\
		\>\textsf{isa}: $\rel{Works\_in}[1,2]$\\
		\textsf{attribute} $\rel{emp\_name}$ \textsf{of} $\rel{Employee}$ \\
		\textsf{attribute} $\rel{dept\_name}$ \textsf{of} $\rel{Dept}$ \\
		\textsf{attribute} $\rel{since}$ \textsf{of} $\rel{Works\_in}$ \\
	\end{tabbing}

	Figure~\ref{fig:ER} depicts $\C$ in the usual graphical notation for the ER model (components are indicated by integers for the relationships).
	The relational schema $\R$ associated with $\C$ consists of the predicates $\rel{manager}/1$, $\rel{employee}/1$, $\rel{dept}/1$, $\rel{works\_in}/2$, $\rel{manages}/2$, $\rel{emp\_name}/2$, $\rel{dept\_name}/2$, $\rel{since}/3$.
	The schema describes employees working in departments of a firm, and managers that are also employees, and manage departments.
	Managers who manage a department also work in the same department, as imposed by the is-a among the two relationships; the permutation $[1,2]$ labeling the arrow denotes that the is-a holds considering the components in the same order (in general, any permutation of $(1,\ldots,n)$ is possible for an is-a between two $n$-ary relationships).
	The constraint $(1,1)$ on the participation of $\eer{Employee}$ in $\eer{Works\_In}$ imposes that every instance of $\eer{Employee}$ participates at least once (mandatory participation) and at most once (functional participation) in $\eer{Works\_In}$; the same constraints hold on the participation of $\eer{Manager}$ in $\eer{Manages}$.
	Suppose we want to know the names of the managers who manage the toy department (named $\const{toy\_dept}$).
	The corresponding conjunctive query over $\C$ is
	$$\begin{array}{rcl}
		q(Z) &\la&
		\rel{manager}(X), \rel{emp\_name}(X,Z), \rel{manages}(X,Y),
		\rel{dept}(Y), \\
		&& \rel{dept\_name}(Y,\const{toy\_dept})\vspace{-.5cm}
	\end{array}$$
\end{example}


The intended semantics of an EER schema is immediately captured by a translation into the relational model that imposes additional constraints to the associated relational schema.
Once we have defined the relational schema $\R$ for an EER schema $\C$, we give the semantics of each construct of the EER model; this is done by specifying what databases (i.e., extensions of the predicates of $\R$) \emph{satisfy} the constraints imposed by the constructs of the EER diagram.
We do that by making use of the relational database constraints introduced in Section~\ref{sec:preliminaries}.
We remind the reader that each entity $E$ in $\C$ has an associated relational predicate $e$ in $\R$, denoted with the same letter, lowercase instead of uppercase; similarly, an attribute $A$ has associated a predicate $a$ and a relationship $R$ a predicate $r$.

\begin{compactenum}[\itshape (1)]
	\item For each attribute $A/2$ for an entity $E$ in an attribute definition in $\C$, we have the ID $a[1] \subseteq e[1]$.
	\item For each attribute $A/(n+1)$ for a relationship $R/n$ in an attribute definition in $\C$, we have the ID $a[1,\ldots,n] \subseteq r[1,\ldots,n]$.
	\item For each relationship $R$ involving an entity $E_i$ as i-th component according to the corresponding relationship definition in $\C$, we have the ID $r[i] \subseteq e_i[1]$.
	\item For each mandatory attribute $A/2$ of an entity $E$ in an attribute definition in $\C$, we have the ID $e[1] \subseteq a[1]$.
	\item For each mandatory attribute $A/(n+1)$ of a relationship $R/n$ in an attribute definition in $\C$, we have the ID $r[1,\ldots,n] \subseteq a[1,\ldots,n]$.
	\item For each functional attribute $A/2$ of an entity $E$ in an attribute definition in $\C$, we have the KD $\key{a}=\{1\}$, since  there cannot be more than one value for attribute $A$ that is assigned to a single instance of $E$.
	\item For each functional attribute $A/(n+1)$ of a relationship $R/n$ in an attribute definition of $\C$, we have the KD $\key{a} = \{1,\ldots,n\}$, since there cannot be more than one value for attribute $A$ that is assigned to a single instance of $R$.
	\item For each is-a relation between entities $E_1$ and $E_2$, in an entity definition in $\C$, we have the ID $e_1[1] \subseteq e_2[1]$, since the is-a relation specifies a set containment between entities $E_1$ and $E_2$.
	\item For each is-a relation between relationships $R_1$ and $R_2$, where components $1,\ldots,n$ of $R_1$ correspond to components $j_1,\ldots,j_n$, in a relationship definition in $\C$, we have the ID: $r_1[1,\ldots,n] \subseteq r_2[j_1,\ldots,j_n]$, since the is-a relation specifies a set containment between relationships $R_1$ and $R_2$.
	\item For each mandatory participation (participation with minimum cardinality $1$) as $c$-th component of an entity $E$ in a relationship $R$, specified by a clause \textsf{participates}$\geq 1$: $R:c$ in an entity definition in $\C$, we have the ID $e[1] \subseteq r[c]$.
	\item For each participation with maximum cardinality $1$ as $c$-th component of an entity $E$ in a relationship $R$, specified by a clause \textsf{participates}$\leq 1$: $R:c$ in an entity definition in $\C$, we have the KD $\key{r} = \{ c \}$.
\end{compactenum}

\begin{definition}[Conceptual dependencies]
	Consider a schema $\R$ and a set of dependencies $\dep=\idep \cup \kdep$, where $\idep$ is a set of inclusion dependencies and $\kdep$ is a set of key dependencies expressed over $\R$.
	We say that $\Sigma$ is a set of \emph{conceptual dependencies} (CDs) if there exists an EER schema $\C$ with associated relational schema $\R$ such that $\Sigma$ is obtained from $\C$ by applying the above points (1)-(11).
\end{definition}

\begin{exampleCont}{\ref{exa:ER-query}}\label{exa:ER-query-contd} Consider again the EER schema shown in
  Figure~\ref{fig:ER}.
	The set of conceptual dependencies associated with the EER schema $\C$ to be imposed on the schema $\R$ consists of the following dependencies.
	$$\begin{array}{rrcll}
		\sigma_1:&\rel{dept\_name}[1]&\subseteq& \rel{dept}[1] &\mbox{ (by rule 1)}\\
		\sigma_2:&\rel{emp\_name}[1]&\subseteq& \rel{employee}[1] &\mbox{ (by rule 1)}\\
		\sigma_3:&\rel{since}[1,2]&\subseteq& \rel{works\_in}[1,2] &\mbox{ (by rule 2)}\\
		\sigma_4:&\rel{works\_in}[1]&\subseteq& \rel{employee}[1] &\mbox{ (by rule 3)}\\
		\sigma_5:&\rel{works\_in}[2]&\subseteq& \rel{dept}[1] &\mbox{ (by rule 3)}\\
		\sigma_6:&\rel{manages}[1]&\subseteq& \rel{manager}[1] &\mbox{ (by rule 3)}\\
		\sigma_7:&\rel{manages}[2]&\subseteq& \rel{dept}[1] &\mbox{ (by rule 3)}\\
		\sigma_8:&\rel{manager}[1]&\subseteq& \rel{employee}[1] &\mbox{ (by rule 8)}\\
		\sigma_9:&\rel{manages}[1,2]&\subseteq& \rel{works\_in}[1,2] &\mbox{ (by rule 9)}\\
		\sigma_{10}:&\rel{employee}[1]&\subseteq& \rel{works\_in}[1] &\mbox{ (by rule 10)}\\
		\sigma_{11}:&\rel{manager}[1]&\subseteq& \rel{manages}[1] &\mbox{ (by rule 10)}\\
		\sigma_{12}:&\key{\rel{works\_in}}&=&\{1\} &\mbox{ (by rule 11)}\\
		\sigma_{13}:&\key{\rel{manages}}&=&\{1\} &\mbox{ (by rule 11)}\vspace{-.5cm}
  \end{array}$$
\end{exampleCont}

Now we characterize the form of relational dependencies resulting from the encoding of EER schemata into relational schemata, the proof of which is straightforward.
\begin{proposition}\label{pro:CDs}
	Consider a schema $\R$ and a set of dependencies $\dep=\idep \cup \kdep$, where $\idep$ is a set of inclusion dependencies and $\kdep$ is a set of key dependencies expressed over $\R$.
	Then, $\Sigma$ is a set of CDs if and only if we can partition $\R$ in three sets $\R_R$, $\R_E$, and $\R_A$ such that the following holds.
	\begin{compactenum}[\itshape (a)]
		\item All predicate symbols in $\R_E$ are unary.
		\item All predicate symbols in $\R_R$ and $\R_A$ have arity at least 2.
		\item The dependencies in $\kdep$ have one of the following forms
		\begin{compactenum}[\itshape (1)]
			\item $\key{r}=\{i\}$, with $1\leq i\leq \arity{r}$, where $r\in\R_R$.
			\item $\key{a}=\{1,\ldots,n\}$, where $a\in\R_A$ and $n=\arity{a}-1$.
		\end{compactenum}
		\item The dependencies in $\idep$ have one of the following forms
		\begin{compactenum}[\itshape (1)]
			\item $e_1[1]\subseteq e_2[1]$, where $\{e_1,e_2\}\subseteq\R_E$.
			\item $e[1]\subseteq r[i]$, where $e\in\R_E$, $r\in\R_R$, and $1\leq i\leq \arity{r}$.
			\item $r[i]\subseteq e[1]$, where $r\in\R_R$, $e\in\R_E$, and $1\leq i\leq \arity{r}$.
			\item $r_1[1,\ldots,k]\subseteq r_2[i_1,\ldots,i_k]$, where $\{r_1,r_2\}\subseteq\R_R$, $\arity{r_1}=\arity{r_2} = k$, and $(i_1,\ldots,i_k)$ is a permutation of $(1,\ldots,k)$.
			\item $a[1]\subseteq e[1]$, where $a\in\R_A$ and $e\in\R_E$.
			\item $a[1,\ldots,n]\subseteq r[1,\ldots,n]$, where $a\in\R_A$, $r\in\R_R$, and $n=\arity{r}=\arity{a}-1$.
			\item $e[1]\subseteq a[1]$, where $e\in\R_E$ and $a\in\R_A$.
			\item $r[1,\ldots,n]\subseteq a[1,\ldots,n]$, where $r\in\R_R$, $a\in\R_A$, and $n=\arity{r}=\arity{a}-1$.
		\end{compactenum}
		\item For every predicate $r\in\R_R$ and for $1\leq i\leq \arity{r}$, there exists an ID $r[i]\subseteq e_i[1]$ in $\idep$ such that $e_i\in\R_E$ and there is no $e'_i\in\R_E$, with $e_i\neq e'_i$, such that $r[i]\subseteq e'_i[1]$ is in $\idep$.
		\item For every predicate $a\in\R_A$, there exists an ID $a[1,\ldots,n]\subseteq p[1,\ldots,n]$ in $\idep$ such that $p\in\R_R\cup \R_E$ and $n=\arity{p}=\arity{a}-1$, and there is no $p'\in\R_R\cup\R_E$, with $p\neq p'$, such that $a[1,\ldots,n]\subseteq p'[1,\ldots,n]$ is in $\idep$.
		\item For every ID $e[1]\subseteq r[i]$ in $\idep$, with $e\in\R_E$, $r\in\R_R$, and $1\leq i\leq \arity{r}$, there is an ID $r[i]\subseteq e[1]$ in $\idep$.
		\item For every ID $r[1,\ldots,n]\subseteq a[1,\ldots,n]$ in $\idep$, with $r\in\R_R$, $a\in\R_A$, and $n=\arity{r}=\arity{a}-1$, there is an ID $a[1,\ldots,n]\subseteq r[1,\ldots,n]$ in $\idep$.
		\item For every ID $e[1]\subseteq a[1]$ in $\idep$, with $e\in\R_E$, $a\in\R_A$, and $\arity{a}=2$, there is an ID $a[1]\subseteq e[1]$ in $\idep$.
	\end{compactenum}
\end{proposition}

Being able to encode EER schemata into relational ones, henceforth we will deal with relational schemata only.

The problem of querying incomplete databases under KDs and IDs is in general undecidable~\cite{Cali03t,CaLR03}.
The largest subclass of functional dependencies\footnote{Functional dependencies are a generalization of key dependencies~\cite{AbHV95}.} and inclusion dependencies for which query answering is known to be decidable is the class of keys and non-key conflicting inclusion dependencies~\cite{Cali03t,CaLR03}.
The main contribution of the present paper is a technique for solving the problem of querying incomplete databases under CDs. This is relevant because EER schemata are very important in practice and CDs are able to capture them.
Our solution consists in a technique for rewriting the given query such that the evaluation of the rewritten query returns the certain answers.

Note that our definition of certain answer, defined in Section~\ref{sec:preliminaries}, considers databases that may also be of infinite size.
In the database literature, interest is typically devoted to databases of finite size only. In particular, the certain answers under finite models can be defined as follows.
\begin{definition}[Certain answer under finite models]\label{def:answers-finite}
	Consider a relational schema $\R$ with a set of dependencies $\dep$, and a finite database $D$ for $\R$.
	Let $q$ be a conjunctive query of arity $n$ over $\R$.
	A $n$-tuple $t$ is a \emph{certain answer under finite models} to $q$ w.r.t.~$D$ and $\dep$ if and only if, for every \emph{finite} database $B$ for $\R$ such that $B \models \dep$ and $B \supseteq D$, we have $t \in q(B)$, and $t$ consists of constants in $\dom$.
	The set of certain answers under finite models is denoted by $\ansf{q}{\dep}{D}$.
\end{definition}
We now show that under CDs, in general, $\ans{q}{\dep}{D}\neq\ansf{q}{\dep}{D}$.
\begin{example}
Consider the following EER schema:
\begin{tabbing}
	\hspace*{1cm}\=\quad\=\+\kill
	\textsf{entity} $\rel{B}$\\
	\>\textsf{participates}($\geq 1$): $\rel{R}:2$\\
	\textsf{entity} $\rel{A}$\\
	\>\textsf{isa}: $\rel{B}$\\
	\>\textsf{participates}($\leq 1$): $\rel{R}:1$\\
	\textsf{relationship} $\rel{R}$ \textsf{among} $\rel{A},\rel{B}$\\
\end{tabbing}
This corresponds to the following set of CDs:
$$
	\Sigma = \left\{
	\begin{array}{rl}
 r[1] \subseteq a[1],\\
 r[2] \subseteq b[1],\\ 
 a[1] \subseteq b[1],\\ 
 b[1] \subseteq r[2],\\ 
 \key{r} = \{1\} 
\end{array}\right.$$
It can be straightforwardly seen that, for every \emph{finite} database $B\supseteq D$ such that $B\models \Sigma$, we have $a(c)\in B$.
Consequently, $\tup{c}\in\ansf{q}{\Sigma}{D}$, where $q$ is the query $q(x) \leftarrow a(x)$.
On the other hand, consider the following database $D_{\infty}$.
$$\begin{array}{rl}
D_{\infty}=\{&\!\!\!\!\! b(c), r(c_1, c), a(c_1), b(c_1), r(c_2, c_1), a(c_2), b(c_2), r(c_3, c_2),\dots\\
&\!\!\!\!\! \dots, a(c_i), b(c_i), r(c_{i+1}, c_i), \dots\}
\end{array}$$
We have that $D_{\infty}\supseteq D$ and $D_{\infty}\models \Sigma$, but $a(c)\not\in D_{\infty}$ and thus 
$\tup{c}\notin\ans{q}{\Sigma}{D}$, therefore we immediately have $\ans{q}{\Sigma}{D}\neq \ansf{q}{\Sigma}{D}$.
\end{example}
Henceforth, we shall not restrict our attention to finite databases only, thus allowing for models of infinite size.

\section{Query Answering with the Chase}
\label{sec:chase}

In this section we introduce the notion of \emph{chase}, which is a fundamental tool for dealing with database constraints~\cite{MaMS79,MaSY81,Vard83,JoKl84}; then we show some relevant properties of the chase under conceptual dependencies (CDs) regarding conjunctive query answering, that will pave the way for the query rewriting technique that will be presented in the next section.

The \emph{chase}%
~\cite{MaMS79,JoKl84} is a key concept in particular in the context of functional and inclusion dependencies.
Intuitively, given a database, its facts in general do not satisfy the dependencies; the idea of the chase is to convert the initial facts into a new set of facts constituting a database that satisfies the dependencies, possibly by collapsing facts (according to KDs) or adding new facts (according to IDs).
When new facts are added, some of the constants need to be \emph{fresh}, as we shall see in the following.
The technique to construct a chase is well known for functional and inclusion dependencies (see, e.g.,~\cite{JoKl84}); however we detail this technique here, since we have adapted it to the simpler case of KDs instead of functional dependencies.

\subsection{Construction of the chase.}
\label{sec:chase-construction}
In order to construct the chase for a database for a relational schema $\R$ with dependencies $\dep=\idep \cup \kdep$, where $\idep$ is a set of inclusion dependencies and $\kdep$ is a set of key dependencies, we use the following rules for IDs and KDs, which apply to a set of facts (i.e., a database instance) and produce a new set of facts.
We indicate as $D$ the set of facts before the application of a rule.


\textsc{Inclusion Dependency Chase Rule.}
Let $r,s$ be relational symbols in $\R$.
Suppose there is a tuple $t$ in
$r^{D}$, and there is an ID $\sigma \in \idep$ of the form $r[\ins{X}_r] \subseteq s[\ins{X}_s]$.
If there is no tuple $t'$ in $s^{D}$ such that $t'[\insX_s]=t[\insX_r]$ (in this case we say the rule is \emph{applicable}), then we add a new tuple $t_{\mathit{chase}}$ in $s^{D}$ such that $t_{\mathit{chase}}[\insX_s]=t[\insX_r]$, and for every attribute $A_i$ of $s$ such that
$A_i \notin \insX_s$, $t_{\mathit{chase}}[A_i]$ is a fresh value in $\freshdom$ that \emph{follows}, according to lexicographic order, all the values already present in the chase.
Note also that we assume that all the values in $\freshdom$ follow, according to lexicographic order, all the values in $\dom$.

\textsc{Key Dependency Chase Rule.}
Let $r$ be a relational symbol in $\R$.
Suppose there is a KD $\kappa$ of the form $\key{r} = \insX$.
If there are two \emph{distinct} tuples
$t,t' \in r^{D}$ such that $t[\insX] = t'[\insX]$ (in this case we say the rule is \emph{applicable}), make the symbols in $t$ and $t'$ equal in the following way.
Let $\insY=\dd{Y}{\ell}$ be the attributes of $r$ that are not in $\insX$; for all $i\in\{1,\ldots,\ell\}$, make $t[Y_i]$ and $t'[Y_i]$ merge into a combined symbol according to the following criterion:
\begin{inparaenum}[\itshape (i)]
	\item if both are constants in $\dom$ and they are not equal, the rule fails to apply and the chase construction process is halted; 
	\item if one is in $\dom$ and the other is a fresh constant in $\freshdom$, let the combined symbol be the non-fresh constant;  
	\item if both are in $\freshdom$, let the combined symbol be the one preceding the other in lexicographic order.
\end{inparaenum}
Finally, replace all occurrences in
$D$ of $t[Y_i]$ and $t'[Y_i]$ with their combined symbol.

Now we come to the formal definition of the chase, which uses the notion of \emph{level} of a tuple; intuitively, the lower the level of a tuple, the earlier the tuple has been constructed in the chase.
In order to make all steps in the construction of the chase univocally determined by the definition, we assume that all facts can be sorted according to lexicographic order (e.g., by using a string comprising the predicate name and the names of all constants in the fact), and so can all pairs of facts as well as all dependencies (e.g., also by using strings that encode them).
%
\begin{definition}[Chase] \label{def:chase}
	Let $D$ be a database for a schema $\R$, and $\dep$ a set of CDs.
	We call \emph{chase} of $D$ according to $\dep$, denoted $\chase{\dep}{D}$, the database constructed from $D$ by  repeatedly executing the following steps, while the KD and ID chase rules are applicable; every tuple $t\in\chase{\dep}{D}$ is also assigned a \emph{level}, denoted by $\level{t}$; if $t\in D$, then $\level{t}=0$.
	\begin{asparaenum}[\itshape (1)]
		\item While there are pairs of facts on which the KD chase rule is applicable, 
		take the pair $t_1,t_2$ such that $min(\level{t_1},\level{t_2})$ is minimal (if there is more than one, take the pair that comes first in lexicographic order) and apply the KD chase rule on $t_1,t_2$ w.r.t. a KD $\kappa$ (if there is more than one KD for which the KD chase rule is applicable on $t_1,t_2$, take the KD that comes first in lexicographic order) so that $t_1,t_2$ collapse into a fact $t_3$; if the rule fails, the chase cannot be constructed and, thus, does not exist; else we define $\level{t_3}=min(\level{t_1},\level{t_2})$.
		\item If there are facts on which the ID chase rule is applicable w.r.t. a full-width ID, choose \emph{the one} (say $t'$) at the lowest level that lexicographically comes first and apply the ID chase rule on $t'$ w.r.t. a full-width ID $\sigma$ (if there is more than one full-width ID for which the ID chase rule is applicable on $t'$, take the full-width ID that comes first in lexicographic order) to generate a new fact $t''$; else, if there are facts on which the ID chase rule is applicable, choose \emph{the one} (say $t'$) at the lowest level that lexicographically comes first and apply the ID chase rule on $t'$ w.r.t. an ID $\sigma$ (if there is more than one ID for which the ID chase rule is applicable on $t'$, take the ID that comes first in lexicographic order) to generate a new fact $t''$.
		We define $\level{t''}=\level{t'}+1$.
	\end{asparaenum}\vspace{-.5cm}
\end{definition}
Note that, according to Definition~\ref{def:chase}, the chase is constructed by applying the KD chase rule as long as possible, then the ID chase rule exactly once, then the KD chase rule as long as possible, etc., until no more rule is applicable. Also, the particular sequence of chase rules to be applied is determined according to a precise lexicographic order, so that there is exactly one chase for a given initial database and set of CDs.
\begin{tobechecked}
 However, a different order would not lead to any significant change in the outcome, since the obtained chase, as discussed in the proof of Theorem~\ref{the:maximum-level}, would be isomorphic with the one obtained in Definition~\ref{def:chase}.
\end{tobechecked}

As we pointed out before, the aim of the construction of the chase is to make the initial database satisfy the KDs and the IDs, by repairing the violations of the constraints.
The obtained (possibly infinite) instance is a representative of all databases that are a superset of the initial database
and satisfy the constraints.
Notice that key dependency violations cannot be repaired by constructing a chase, but would require an explicit treatment, as explained in Section~\ref{sec:extensions}; in such a case the chase does not exist.
It is easy to see that $\chase{\dep}{D}$ can be infinite only if the set of IDs in $\dep$ is \emph{cyclic}~\cite{AbHV95,JoKl84}, i.e., if there is a sequence of IDs in $\dep$ of the form $r_1[\insX_1]\subseteq r_2[\insX_1'], r_2[\insX_2]\subseteq r_3[\insX_2'], \ldots, r_n[\insX_n]\subseteq r_{n+1}[\insX_n']$ and $r_{n+1}=r_1$.
In the following we will show how the chase can be used in computing the answers to queries over incomplete databases under dependencies.

\subsection{Query Answering and the Chase.}
\label{sec:chase-answering}
In their milestone paper~\cite{JoKl84}, Johnson and Klug proved that, under certain subclasses of KDs and IDs, a containment between two conjunctive queries $q_1$ and $q_2$ can be tested by verifying the existence of a so-called query homomorphism.
Roughly speaking, such a homomorphism has to map the body of $q_2$ to the chase of the body of $q_1$, and the head of $q_2$ to the head of $q_1$.
Johnson and Klug proved that, in order to test containment of CQs under IDs alone or \emph{key-based} dependencies (a special class of KDs and IDs), it is sufficient to consider a \emph{finite}, initial portion of the chase.
The result of~\cite{JoKl84} was extended in~\cite{CaLR03} to a broader class of dependencies, strictly more general than keys with foreign keys: the class of KDs and \emph{non-key-conflicting inclusion dependencies (NKCIDs)}~\cite{Cali03t}, that behave like IDs alone because NKCIDs do not interfere with KDs in the construction of the chase.
The above results about query containment (see, e.g.,~\cite{CaGK08}) can be straightforwardly adapted to solve the decision problem of answering on incomplete databases, since, as it will be shown later, the chase is a representative of all databases that satisfy the dependencies and are a superset of the initial data.

In a set of CDs, IDs are not non-key-conflicting (or better \emph{key-conflicting}), therefore the decidability of query answering cannot be deduced from~\cite{JoKl84,CaLR03}, (though it can be derived from~\cite{CaDL98}, as we shall discuss later).
In particular, under CDs, the construction of the chase has to face interactions between KDs and IDs; this can be seen in the following example, taken from~\cite{Cali06}.

\begin{example} \label{exa:chase} Consider again the EER schema of
  Example~\ref{exa:ER-query}.  Suppose we have an initial (incomplete)
  database, with the facts $\rel{manager}(m)$ and $\rel{works\_in}(m,d)$.  If
  we construct the chase, we obtain the facts $\rel{employee}(m)$,
  $\rel{manages}(m,\alpha_1)$, $\rel{works\_in}(m,\alpha_1)$,
  $\rel{dept}(\alpha_1)$, where $\alpha_1$ is a fresh constant.  Observe that
  $m$ cannot participate more than once in $\rel{works\_in}$, so we deduce
  $\alpha_1 = d$.  We must therefore replace $\alpha_1$ with $d$ in the rest
  of the chase, including the part that has been constructed so far.
Therefore, $\chase{\dep}{D}=\{\rel{manager}(m), \rel{works\_in}(m,d), \rel{employee}(m), \rel{manages}(m,d), \rel{dept}(d)\}$.
\end{example}
In spite of the potentially harmful interaction between IDs and KDs,
analogously to the case of IDs alone~\cite{CCDL04}, it can be proved that, in
the presence of CDs, the chase is a representative of all databases that are a
superset of the initial (incomplete) data, and satisfy the dependencies;
therefore, it serves as a tool for query answering, as shown in
Theorem~\ref{the:answering-on-chase} below.


As was made explicit in Definition~\ref{def:chase}, the chase may not exist if
some application of the KD rule fails.  This may happen even when the database
satisfies the key dependencies, as shown in the next example.

\begin{example} \label{exa:failing-chase} Consider two binary predicates $r$
  and $s$, derived from two binary relationships $R$ and $S$, for which there
  is an is-a relation (that generates the ID $r[1,2]\subseteq s[1,2]$) and a
  participation with maximum cardinality $1$ for the first component of $s$
  (that generates the KD $\key{S}=\{1\}$). The mentioned ID and KD are a fragment of a set of CDs that is sufficient to show that the chase may not exist even if the initial database satisfies the dependencies. Let the initial database be
  $D=\{r(a,b), s(a,c)\}$. Although $D$ satisfies the KD, the chase rule for
  the ID generates a tuple $s(a,b)$, which triggers a (failing) KD chase rule
  application on $s(a,b)$ and $s(a,c)$. Therefore the chase for this database
  and constraints does not exist.
\end{example}
Since the chase may be of infinite size, it would seem that checking whether a
chase exists is semi-decidable.  Indeed, in the general case of IDs and KDs it
is not known whether it is decidable to check whether the chase exists.  

However, the following lemma shows that 
termination of the chase under CDs is decidable;
we will
then use it to state some of our results.

\begin{lemma} \label{lem:chase-exists-decidable} Let $D$ be a database for a
  relational schema $\R$ and $\dep$ a set of CDs over $\R$.  Then, checking
  whether $\chase{\dep}{D}$ exists is decidable in time polynomial in the size
  of $D$.
\end{lemma}
\begin{proof}
  We start by observing that the application of a \emph{unary} ID (i.e., an ID
  that involves a single attribute) cannot cause a failure of the chase by
  violation of a KD: indeed, considering a generic unary ID $r_1[k_1]
  \subseteq r_2[k_2]$, the only possible violation of a KD due to the
  application of this ID is when we have the KD $\key{r_2} = \{k_2\}$;
  however, such violation never causes a failure of the chase, since all
  values in the added tuple that are in positions different from $k_2$ are all
  fresh constants.  Now, let us indicate with $\rdep$ the IDs in $\dep$ that
  derive from is-a relations among relationships; they are IDs of the form
  $r_1[1,\ldots,n] \subseteq r_2[j_1,\ldots,j_n]$, where $j_1,\ldots,j_n$ is a
  permutation of $1, \ldots n$ and both $r_1$ and $r_2$ have arity $n$.  It is
  immediately seen that:
\begin{compactenum}[\itshape (i)]
  \item Facts in the chase of the form $r(c_1, \ldots, c_n)$, where $r$ is a
    relation belonging to the set $\R_R$ of $n$-ary relationships in the
    conceptual schema, contain
\begin{compactitem}
	\item only non-fresh constants, 
	\item only fresh constants, or
	\item exactly one non-fresh constant (possibly occurring more than once).
\end{compactitem}
No other case is possible.
This can be shown by induction on the number of application of chase rules. Consider also that
\begin{compactitem}
	\item Facts regarding (unary) predicates associated with entities may either contain a fresh or a non-fresh constant. 
	\item For facts regarding predicates associated with $n$-ary attributes, we have that the last position may be occupied by either a fresh or a non-fresh constant, and the first $n$ positions behave like a fact for a relation in $\R_R$ (i.e., they contain only non-fresh constants, only fresh constants, or exactly one non-fresh constant).
\end{compactitem}
In the base case (no application), we only have facts in $D$, which only contain non-fresh constants. Suppose now, by inductive hypothesis, that, after $i$ applications of the chase rules, the facts are only of the forms mentioned above. The inductive step consists in showing that no new application of a chase rule produces facts that are not in one of the forms mentioned above. To see this, it suffices to verify this for all forms (1)-(11) of dependencies that may occur in CDs, as described in Section~\ref{sec:chase}. This is immediate for (1)-(10). As for (11), consider that a KD rule can be applied on two tuples $t_1$ and $t_2$ for a relation $r\in\R_R$ in two cases:
\begin{compactitem}
	\item $t_1$ and $t_2$ both have in the position of the key the same non-fresh constant. In this case the inductive step immediately follows, by either a failure of the chase or the generation of a new tuple containing exactly one non-fresh constant (possibly occurring more than once).
	\item $t_1$ and $t_2$ both have in the position of the key the same fresh constant. The inductive step follows immediately, unless $t_1$ contains exactly one non-fresh constant, say $c$, and $t_2$ contains exactly one non-fresh constant, say $d$, with $d \neq c$, because then the KD rule could produce a tuple containing two different non-fresh constants. However, this case cannot occur. To see this, it suffices to show that if two tuples $t_1$ and $t_2$ for $r\in\R_R$ have a fresh constant in common, then they cannot have different non-fresh constants.
	This can, again, be shown by induction on the applications of chase rules for dependencies of the forms (1)-(11). Basically, the only way for tuples of relations in $\R_R$ to have fresh constants in common is to apply chase rules on dependencies of the forms (9)-(11).
	\begin{compactitem}
		\item With form (9), the application of the ID chase rule on a cycle of is-a relations between relationships may generate two tuples sharing a fresh constant. However, only permutations of the positions can take place, but the constants are unchanged.
		\item Two applications of the ID chase rule on two different IDs of form (10) for the same entity and the same relationship but on two different components can generate two tuples sharing a fresh constant. However, all the other constants will also be fresh.
		\item The application of a KD chase rule for a KD of form (11) is now trivially harmless by inductive hypothesis.
	\end{compactitem}
\end{compactitem}
  \item All facts of the form $r(c_1, \ldots, c_n)$, with $r \in \R_R$,
    that contain only non-fresh constants are obtained by applying (possibly several times) the ID chase rule for IDs in
    $\rdep$ to facts in the initial database (constituted in turn by tuples containing
    only non-fresh constants).
  \item By what stated in point (i) above, the only way of causing a failure in the chase construction (apart from violations of key constraints already in $D$) is to
    apply an ID in $\rdep$ to a tuple having only non-fresh constants, thus
    introducing a (non-repairable) violation of some KD due to the presence of
    another tuple having only non-fresh constants; in all other cases, every violation of a
    KD is repaired by applications of the KD chase rule.
\end{compactenum}
  This said, it follows that if there is no failure in $\chase{\rdep}{D}$,
  there is no failure in $\chase{\dep}{D}$.  It remains to check whether
  $\chase{\rdep}{D}$ is finite: it is easily seen that it indeed cannot be infinite, since every tuple in
  $\chase{\rdep}{D}$ is of the form $r(c_1, \ldots, c_n)$, with $r \in \R_R$,
  and where $c_1, \ldots, c_n$ are obtained by a permutation of $d_1, \ldots,
  d_n$, where the fact $r'(d_1, \ldots, d_n)$, with $r' \in \R_R$, is in the
  initial database $D$.  The maximum depth of $\chase{\rdep}{D}$ is $W!$,
  where $W$ is the maximum arity of predicates in $\R$.
  It is also straightforward to see that the size of $\chase{\rdep}{D}$ is
  polynomial in $|D|$ (size of $D$, i.e., number of tuples of $D$), and that
  $\chase{\rdep}{D}$ can be constructed in time polynomial in $|D|$.
  By the above considerations, it is immediately seen that $\chase{\rdep}{D}$ fails iff $\chase{\dep}{D}$ fails.
  The thesis follows.
\end{proof}

\begin{lemma} \label{lem:chase-satisfies-deps} Let $D$ be a database for a
  relational schema $\R$ and $\dep$ a set of CDs over $\R$ such that
  $\chase{\dep}{D}$ exists.  Then $\chase{\dep}{D}\models \dep$.
\end{lemma}

\begin{proof}
  Trivial, by the construction of Definition~\ref{def:chase}.
\end{proof}

The following lemma is a technical result that will be used in the proof of
Theorem~\ref{the:answering-on-chase}.  Informally, it shows that the chase of
a database $D$, when it exists, is a powerful tool for answering queries: for
every solution $B$ (database that is a superset of the given incomplete
database $D$ and that satisfies the constraints), there is a homomorphism that
sends the chase of $D$ onto $B$.
This result follows from the results in~\cite{FKMP05,DNR08}, but we provide a direct proof for the sake of completeness.

\begin{lemma} \label{lem:homomorphism} Let $D$ be a database for a relational
  schema $\R$ and $\dep$ a set of CDs over $\R$ such that $\chase{\dep}{D}$
  exists.
  Then, for every database $B$ for $\R$ such that $B \models \dep$ and $B
  \supseteq D$, we have that there exists a homomorphism 
from $\chase{\dep}{D}$ to $B$.
\end{lemma}

\begin{proof}
  Similarly to what is done for the analogous result in~\cite{CCDL04}, we proceed by induction on the applications of the (ID or KD) chase rules.
We define a homomorphism
  $\mu$ inductively, and we simultaneously show that for each relation $r$ of
  arity $n$ in $\R$, and each tuple $(c_1, \dots, c_n)$ constituted by
  elements in $\dom\cup\freshdom$, if $(c_1, \dots, c_n)\in
  r^{\chase{\dep}{D}}$, then $(\mu(c_1), \dots, \mu(c_n))\in r^{B}$.
\begin{asparaenum}[\itshape (1)]
	  \item \textbf{Base case}.
		After $0$ applications of a chase rule, the constructed part of the chase coincides with $D$.
		Since $B\supseteq D$, the mapping $\mu$ that maps each constant in $D$ into itself is a homomorphism from the constructed part of the chase to $B$.

	  \item \textbf{Inductive step}.
		\emph{First case: the applied rule is the ID chase rule.}
		Suppose that in the application of the rule, we are inserting the tuple $t^{*}=(\dd{\alpha}{n})$ in $\chase{\dep}{D}$, where $r$ has arity $n$, $\alpha_i\in\freshdom$ for each $i\neq k$, $\alpha_k\in\dom\cup\freshdom$, and the tuple is inserted in $r^{\chase{\dep}{D}}$ because of the ID $w[j] \subseteq r[k]$ (other forms of IDs among those described in points (1)-(11) in Section~\ref{sec:conceptual-model} are dealt with similarly).
		Since we are applying the rule because of the dependency $w[j] \subseteq r[k]$, there is a tuple $t$ in $w^{\chase{\dep}{D}}$ such that $t[j] = \alpha_k$.
		By inductive hypothesis, there is a constant $c_k$ in $\dom$ such that $\mu(\alpha_k) = c_k$, and there is a tuple $t'\in w^{B}$ such that for each $i$, $t'[i] = \mu(t[i])$, with $t'[j] =\mu(\alpha_k) = c_k$.
		Because of the constraint $w[j] \subseteq r[k]$, and because $B$ satisfies the constraints, there is a tuple $t''$ in $r^B$ with $t''[k]=c_k$; let then $t''=(\dd{c}{n})$.
		Then, we set $\mu(\alpha_i) = c_i$ for each $i\neq k$, and we can conclude that
		$\mu(t^{*})\in r^B$.

		\emph{Second case: the applied rule is the KD chase rule.}
		  By inductive hypothesis, there exists a homomorphism $\mu$ mapping the two tuples $t,t'$ on which the KD rule is applied into tuples $\mu(t)$ and $\mu(t')$ in $B$. Note that, since the KD rule is applicable to $t,t'$ and $B\models \dep$, we must have $\mu(t)=\mu(t')$.
		In the chase, $t$ and $t'$ are then replaced by a new tuple, say $t''$, that contains (in the same positions) all the non-fresh constants of $t,t'$ and a subset of the fresh constants of $t,t'$ (some of which may disappear by the KD chase rule), but no new fresh constant.
		Therefore, $\mu$ trivially also maps $t''$, as well as all other tuples in the chase, into facts of $B$.
\end{asparaenum}
\end{proof}

The following theorem is the main result of this section, and it characterizes
the chase as a formal tool for query answering under KDs and IDs.  In
particular, the theorem states that the answers to a query $q$, posed on an
incomplete database $D$ under a set $\dep$ of CDs, can be obtained by
evaluating $q$ over the chase of $D$ w.r.t.{} $\dep$, $\chase{\dep}{D}$, and
discarding the result tuples that contain at least one fresh value.

\begin{theorem} \label{the:answering-on-chase} Let $D$ be a database for a
  relational schema $\R$ and $\dep$ a set of CDs over $\R$ such that
  $\chase{\dep}{D}$ exists.  Then, for every conjunctive query $q$ over $\R$,
  we have that $\nofresh{q}(\chase{\dep}{D}) = \ans{q}{\dep}{D}$.
\end{theorem}

\begin{proof}
  The theorem is proved by considering a generic database $B$ such that $B
  \models \dep$ and $B \supseteq D$.
	
  By Lemma~\ref{lem:homomorphism} we derive the existence of a homomorphism
  $\mu$ that sends the facts of $\chase{\dep}{D}$ to facts of $B$; if $t \in
  q(\chase{\dep}{D})$, there is a homomorphism $\lambda$ from the atoms of
  $\body{q}$ to $\chase{\dep}{D}$ that sends $\head{q}$ to $t$; therefore, the
  composition $\lambda \circ \mu$ is a homomorphism from the atoms of
  $\body{q}$ to $B$ that sends $\head{q}$ to $t$, which proves
  $q(\chase{\dep}{D}) \subseteq \ans{q}{\dep}{D}$, and, a fortiori,
  $\nofresh{q}(\chase{\dep}{D})\subseteq \ans{q}{\dep}{D}$.
	
  For the other inclusion, consider that $\chase{\dep}{D} \supseteq D$ and
  $\chase{\dep}{D} \models \dep$. Then, by Definition~\ref{def:answers} we
  have that a tuple $t$ is a certain answer to $q$ in $D$ under $\dep$ only if
  it is an answer to $q$ in $\chase{\dep}{D}$ with no fresh constant; hence
  $\nofresh{q}(\chase{\dep}{D}) \supseteq \ans{q}{\dep}{D}$.
\end{proof}

Notice that Theorem~\ref{the:answering-on-chase} does not lead to an algorithm
for query answering (apart from special cases), since the chase may have
infinite size.


\section{Answering Queries by Rewriting}
\label{sec:answering}

In this section we present an efficient technique for query answering
on incomplete data in the presence of CDs; such technique is based on
\emph{query rewriting}; in particular, the answers to a query are obtained by
evaluating a new query, obtained by rewriting the original one according
to the dependencies, over the initial incomplete data.

For the sake of simplicity, in the remainder of this section we shall disregard attributes from our treatment, since attributes are acyclic and therefore can be added without changing the results.

\subsection{Query rewriting}
%
Query answering under CDs can be decided by checking an initial segment of the chase of a database.
We show that the certain answers to a CQ $q$ over a database $D$ can be computed by evaluating $q$ over the initial segment of the chase of $D$, whose size, defined by a maximum level $\maxlevel$, depends on the query, on the dependencies, and on the size $\sizeConnected$ of the largest connected part of the join
graph of database $D$. The 
join graph of a database $D$ is an undirected graph that has as nodes
 the atoms of $D$ and has an arc $(\atom{A},\atom{B})$ iff $\atom{A}$ and $\atom{B}$ share a constant.

\newcommand{\maxlevelvalue}[0]{ 4\cdot|q| \cdot \deltavalue}
\newcommand{\maxlevelvaluedelta}[0]{4\cdot|q|\cdot\delta}
\newcommand{\deltavalue}[0]{|\R|\cdot (W+1)^W}
\begin{theorem}\label{the:maximum-level}
	Let $\R$ be a relational schema, $\dep$ a set of CDs over $\R$, $q$ a conjunctive query over $\R$, and $D$ a database for which $\chase{\dep}{D}$ exists.
	Then, there is a number $\maxlevel$ that depends on $q$, $\dep$, $\R$, and $\sizeConnected$ such that for every tuple $t\in \nofresh{q}(\chase{\dep}{D})$, there exists a homomorphism $\mu$ sending $\body{q}$ to facts of $\chase{\dep}{D}$ and $\head{q}$ to $t$ such that all the atoms in $\mu(\body{q})$ are in the first $\maxlevel$ levels of $\chase{\dep}{D}$.
	%
\end{theorem}

\begin{proof}
First of all, we introduce the \emph{chase forest} for $\chase{\dep}{D}$ given a database $D$ and a set of CDs $\dep$.
The nodes of the forest are the atoms in $\chase{\dep}{D}$, and there is an arc $(\atom{A}_1, \atom{A}_2)$ iff $\atom{A}_2$ is generated from $\atom{A}_1$ by an application of the ID chase rule.
The roots in the forest are the atoms in $D$, and they are at level $0$.
If there is an arc $(\atom{A}_1, \atom{A}_2)$ and $\atom{A}_1$ is at level $\ell$, then $\atom{A}_2$ is at level $\ell+1$.
\begin{unused}
	Given a chase forest $F$, we denote with $\subtree{\atom{A}}{F}$ the subtree of $F$ rooted in $\atom{A}$.
	Given a set of constants $S$, we say that two atoms $\atom{A}_1, \atom{A}_2$ are $S$-isomorphic, denoted $\atom{A}_1 \iso_S \atom{A}_2$, iff there exists an isomorphism $\mu$ such that $\mu(\atom{A}_1) = \atom{A}_2$, and both $\mu$ and $\mu^{-1}$ leave the non-fresh constants in $S$ unaltered.
	This notion naturally extends to sets of atoms, trees and forests having atoms as nodes, and paths in such forests.
	Let us also divide the descendants of an atom in equivalence classes called \emph{types}.
	Let each type of an atom $\atom{A}$ consist of atoms that are mutually $\adom{\atom{A}}$-isomorphic, where $\adom{\atom{A}}$ denotes the so-called \emph{active domain} of $\atom{A}$, i.e., the set of (fresh and non-fresh) constants appearing in $\atom{A}$ as arguments; in other words, each type identifies an element of the quotient set $\chase{\dep}{D}/$$\iso_{\adom{\atom{A}}}$.
	We shall say that a given type $\tau$ \emph{occurs} in a set of atoms $\A$ if there is at least one atom of type $\tau$ occurring in $\A$.
	All types that can occur among the descendants of a given atom $\atom{A}$ necessarily already occur within a distance of $\delta$ levels from $\atom{A}$, with $\delta = \deltavalue$. This is due to the fact that $\delta$ is the maximum cardinality possible for the set of types of $\atom{A}$, since a type is obtained by arranging $W+1$ values (at most $W$ in $\adom{\atom{A}}$ plus a ``don't care'' value representing any value not in $\adom{\atom{A}}$) in the positions of an atom (which are at most $W$), once per each relation in $\R$.
	We now prove auxiliary results showing that a constant can be propagated in the chase for at most a fixed number of levels that does not depend on $D$.
	The first lemma, below, regards applications of IDs; the second lemma regards applications of KDs.
\end{unused}
	In order to carry on the proof, we now prove that a constant can be propagated in the chase for at most a fixed number of levels that does not depend on $D$.
\begin{lemma}\label{lem:chase-constants-propagation}
	Let $D$ be a database for a relational schema $\R$, $\dep$ a set of CDs over $\R$ such that $\chase{\dep}{D}$ exists, and $q$ a conjunctive query over $\R$.
	Let $a$ be a constant in $\dom$ occurring in an atom in $D$.
	Then $a$ never occurs in any fact with level greater than $\delta_D = \delta_C\cdot \sizeConnected$ in $\chase{\dep}{D}$, where
	$\delta_C = |\R|\cdot (1 + |\R|\cdot W!)$.
\end{lemma}
\begin{proof}
	We start by considering the IDs.
	First, observe that, in a set of CDs, the only non-unary IDs in $\dep$ are the IDs encoding is-a arcs between relationships (which are full-width IDs) and the IDs regarding attributes of a relationship.
	Clearly, $a$ can be propagated to other atoms by applications of an ID chase rule, starting from the atom $\atom{\theta}\in D$ in which it occurs, then from the atom generated from $\atom{\theta}$ by the application, and so forth. The propagation can be done for up to $|\dep|$ more levels if there are no cycles in the IDs, but also for more, if there are cycles.

	Whenever there is an application of an $n$-ary ID ($n\geq 2$) on an atom $\atom{A}$, the generated atom $\atom{A}'$ contains a permutation of the constants occurring in $\atom{A}$; both the involved predicates have the same arity $n$ (except in the case of an ID regarding attributes of a relationship, where one predicate has arity $n+1$, but the $(n+1)$-th argument is never used in the IDs).
	Then, a sequence of consecutive applications of $n$-ary IDs can go on for at most $n!\cdot|\R|$ levels, since there are $n!$ possible permutations of the constants in $\atom{A}$ and there are at most $|\R|$ relations involved in $n$-ary IDs. All constants occurring in $\atom{A}$ (except at most the last one, if $\atom{A}$ regards an attribute of a relationship) are propagated throughout the sequence.
	
	All other applications regard unary IDs.
	At least one of the two  predicates involved in a unary ID must be unary, and the only way to retain $a$ in a unary atom is that it be of the form $e(a)$, where $e$ is a unary predicate; clearly such fact can be generated only once in the chase, and there are at most $|\R|$ unary predicates in $\R$.

	Any path in the chase starting from $\atom{\theta}$ consists of sequences of consecutive applications of $n$-ary IDs ($n\geq 2$) interleaved by applications of unary IDs.
	According to the previous considerations,
there can be at most $|\R|+1$ sequences of consecutive applications of $n$-ary IDs (with $n\geq 2$ and $n\leq W$).
	Given the maximum lengths of such sequences, $a$ can be propagated for at most 
	 $\delta_C = |\R|\cdot (1 + |\R|\cdot W!)$.

	We now consider the KDs. To prove the claim, we first state
the following lemma.
\begin{lemma}\label{lem:adjacency}
	Let $\atom{A}$ be the first atom (of the form $r(\ldots,z_0,\ldots)$, where $r$ is $n$-ary, $n\geq 2$) in which a constant $z_0\in \dom\cup\freshdom$ occurs, with $\ell=\level{\atom{A}}>\delta_C$.
	Let $\atom{B}$ be the closest predecessor of atom $\atom{A}$ of the form $e(w_0)$ ($e$ unary).
	Let $\atom{B}'$ be an atom of the form $e(z_1)$, $z_1\in \dom\cup\freshdom$, with $\level{\atom{B}'}>\ell+\delta_C$ such that there is an atom $\atom{C}$ of the form $e'(z_0)$ ($e'$ unary) in the path between $\atom{A}$ and $\atom{B}'$.
	Then no constant occurring in $\atom{A}$ other than $z_0$ occurs in any of the descendants of $\atom{B}'$.
\end{lemma}
\begin{proof}
	%
Atom $\atom{C}$ may well have a child (or a descendant obtained by consecutive applications of the ID chase rule for non-unary IDs from the child) $\atom{D}$ of the form $r'(\ldots,z_1,\ldots)$ such that it agrees on the key of $r'$ (on value $z_1$) with some descendant $\atom{D}'$ of $\atom{B}'$ of the same form, so that the constants in $\atom{D}$ (possibly including $z_0$) will replace the corresponding constants of $\atom{D}'$ in all the descendants of $\atom{B}'$.
	Note that $\atom{B}'$ is necessarily a descendant of $\atom{C}$ with the same constants as $\atom{D}$.
	This shows that $z_0$ may well occur in some descendant of $\atom{B}'$.
	Let us indicate with $z'_0$ the constant that is replaced by $z_0$ after the application of the KD chase rule.
	Assume, by contradiction, that one of the constants in $\atom{A}$ other than $z_0$ occurs in some descendant of $\atom{B}'$. Then, there must be a descendant $\atom{A}'$ of $\atom{B}'$ of the form $r(\ldots, z'_0, \ldots)$ that, once $z'_0$ is replaced by $z_0$, fires the application of a KD chase rule between $\atom{A}$ and $\atom{A}'$.
	There are two cases:
	\begin{inparaenum}[\itshape (i)]
		\item $\atom{A}'$ generates  $\atom{D}'$ via a sequence of non-unary IDs. Then $z_1$ is replaced by $w_0$, then the subtree rooted in $\atom{B}'$ gets to have the same root as the subtree rooted in $\atom{B}$ and therefore it disappears as a consequence of the KD application. 
		\item $\atom{A}'$ is a descendant of $\atom{C}'$ along a path that contains at least an application of the ID chase rule for a unary ID, where $\atom{C}'$ is obtained from $\atom{B}'$ by the same sequence of applications of ID chase rules as those generating $\atom{C}$ from $\atom{B}$. Again, the KD chase rule makes $\atom{C}'$ become equal to $\atom{C}$, therefore the whole subtree rooted in $\atom{C}'$ disappears, as easily seen, as above.
	\end{inparaenum}
	%
	%
\end{proof}
Consider the proof of Lemma~\ref{lem:adjacency} and assume $z_0\in\dom$. Then, after at most $\delta_C$ levels $z_0$ will not appear together with any of the other constants in $\atom{A}$. Also, $z_0$ cannot be propagated indefinitely in the chase by applications of ID chase rules, since this requires using $z_0$ with a unary predicate, which can be done only once per unary predicate.
However, if $z_0$ appears in an atom in $D$ together with another constant $c$, then $c$ could appear together with $z_0$ in a descendant of $\atom{B}'$, and propagate through further $\delta_C$ levels. By the same principle, this can go on for every sequence of constants $c_1,\ldots,c_n$ such that $c_i$ occurs in the same atom in $D$ together with $c_{i+1}$.
Since the maximum sequence of this kind can have length $|\sizeConnected|$, and the sequences in $D$ are not altered by the chase construction, the claim follows.
\end{proof}

Lemma~\ref{lem:chase-constants-propagation} is the key property for stopping the construction of the chase at a given level $\maxlevel$ 
without altering query answering.
We first prove the claim for the simple but important subclass of conjunctive queries called non-boolean (i.e., with at least one distinguished variable) connected queries.
A set of atoms $\N$ is \emph{connected} if the undirected graph $(\N,\A)$ is connected, where $\N$ is the set of nodes, and $\A$ is the set containing exactly all arcs between any two atoms in $\N$ that share a variable or a constant.
A CQ $q$ is \emph{connected} if $\body{q}$ is.
Every maximal subset of $\body{q}$ that is connected is called a \emph{connected part} of $q$.
Assume $\mu$ is a homomorphism sending $\head{q}$ to a non-empty tuple $t$ of constants in $\dom$ and $\body{q}$ to atoms of $\chase{\dep}{D}$. Since the query has at least one distinguished variable, then there is at least one atom $\atom{A}$ in $\body{q}$ such that $\mu(\atom{A})$ contains a constant $c_1$ of $t$, that then is in $\dom$. By Lemma~\ref{lem:chase-constants-propagation}, the constants in $\dom$ cannot occur at levels greater than $\delta_D$; then $\level{\mu(\atom{A})}\leq\delta_D$.
If a query is connected and non-boolean, then among the other body atoms there is at least another atom $\atom{A}'$ sharing a variable with $\atom{A}$, and thus such that $\mu(\atom{A}')$ shares a constant with $\mu(\atom{A})$.
Note now that $\mu(\atom{A})$ contains $c_1$ plus possibly other constants.
If such constants are in $\dom$, then also $\mu(\atom{A}')$ has a level at most $\delta_D$.
Else, they are all fresh and have been created in the subtree rooted in the closest unary predecessor $\atom{B}$ of $\mu(\atom{A})$; $\atom{B}$ has the form $e_1(c_1)$.
Now we show that all the constants different from $c_1$ (say, $z_1, \ldots, z_n$) in $\mu{\atom{A}}$ occur within the first $\delta_C$ levels of $\mu{\atom{A}}$, and therefore $\mu(\atom{A}')$ also occurs at a level at most $\level{\mu{\atom{A}}}+\delta_C$.
To see this, we simply reapply Lemma~\ref{lem:adjacency} by considering $\mu{\atom{A}}$ alone as the starting database for the subsequent propagation of constants.
Indeed, for $1\leq i,j\leq n$, the longest path from an atom containing $z_i$ (but not $z_j$) to an atom containing $z_j$ (but not $z_i$) in the join graph is $1$.
This process can be iterated for all the remaining atoms in the query.
Since the size of the longest path in the graph of $q$ is $|q|$, it follows that all the images of the atoms of the query are in the first 
$\delta_M=\delta_D + \delta_C\cdot(|q|-1)$ levels.

If the query is not connected, but each connected part is non-boolean, the same argument as before applies to each connected part, with the same final $\delta_M$.

If the query has at least a boolean connected part, we can reason as follows.
Let $A$ be the atom in the connected part whose image $\mu(\atom{A})$ is at the lowest level among the query atoms.
If $\level{\mu(\atom{A})}>\delta_D$, then there is another homomorphism $\mu'$ sending $\body{q}$ to atoms of $\chase{\dep}{D}$ such that $\level{\mu'(\atom{A})}\leq\delta_D$, because all types occur within the first $\delta_D$ levels, where two atoms have the same type if they share the same predicate and agree on all the positions where a constant of $\dom$ appears. With the same argument as before, all the images via $\mu'$ are at a level at most $\delta_M$.
\end{proof}

\begin{unused}
	\begin{figure}[tb]
	 \centering
	 \resizebox{!}{.24\textheight}{\input{proof.pstex_t}}
	 \caption{Figure for the proof of Theorem~\ref{the:maximum-level} representing $\subtree{\atom{A}_1}{F}$.}
	 \label{fig:maximum-level}
	\end{figure}
	%
\end{unused}

The previous theorem suggests a naive strategy for query answering: first,
compute the initial segment of $\chase{\dep}{D}$, i.e., its first $\maxlevel$
levels, and then evaluate the query $q$ on such a segment. To do that, we also need the following Lemma.
\begin{lemma}\label{lem:back-propagation}
	Consider the application of a KD chase rule on two atoms $\atom{A_1}$ and $\atom{A_2}$ with $\level{\atom{A_1}}=\ell_1>\delta_D$ and $\level{\atom{A_2}}=\ell_2>\delta_D$.
	Consider also  all subsequent applications of KD chase rules before the next application of an ID chase rule.
	Then, after all these applications, no atom in the chase is affected that has level lower than $\min\{\ell_1,\ell_2\}-\delta_C$.
\end{lemma}
\begin{proof}
	By definition of the chase, when a KD chase rule is applied, the affected constants are the more recent ones in the chase construction.
	Then, it easily follows that they may only occur at most $\delta_C$ levels before $\min\{\ell_1,\ell_2\}$.
	Indeed, $\atom{A_1}$ and $\atom{A_2}$ have at least a constant in common. Two cases are possible:
	\begin{inparaenum}[\itshape (i)]
		\item they share a constant in $\dom$, therefore they may only occur within the first $\delta_D$ levels by Lemma~\ref{lem:adjacency}, against the hypotheses;
		\item they share a constant in $\freshdom$.
	\end{inparaenum}
	In the latter case, they have a common unary predecessor $\atom{A_0}$ within $\delta_C$ levels before $\min\{\ell_1,\ell_2\}$.
In this case, the replacement of constants has an impact only on the subtree $T$ rooted in $\atom{A_0}$ since all other constants in $T$ are by construction newer than the one occurring in $\atom{A_0}$.
Ditto for the subsequent applications.
\end{proof}

By Lemma~\ref{lem:back-propagation}, it is immediate to see that the application of the KD chase rule does not affect any facts whose depth is smaller by at least $\delta_C$ levels than the level of the facts involved in the KD; therefore, to compute the first $\maxlevel$ levels of $\chase{\dep}{D}$ means to apply the chase rules of Definition~\ref{def:chase} until no chase rule is applicable on facts at a level smaller than $\maxlevel+\delta_C$.
  However, it is
easy to see that such a strategy would not be efficient in real-world cases,
where $D$ has a large size.  Our plan of attack is then to rewrite $q$
according to the CDs on the schema and on the size $\sizeConnected$ of the largest connected part of the join graph, and then to evaluate the rewritten query over the initial data.
This turns out to be more efficient in practice, if $\sizeConnected$ is bounded or known to be reasonably small, since
it does not involve the entire database $D$ in the query processing, except for the
last evaluation step, so most of the computation is kept at the intensional
level.
%
In particular, the rewritten query is expressed in Datalog, and it is the
union of two sets of rules, denoted $\progid$ and $\progkd$, that take into
account IDs and KDs respectively, plus a set of rules $\progeq$ that simulates
equality.  Finally, function symbols present in the rules will be eliminated
to obtain a Datalog rewriting.

Consider a relational schema $\R$ with a set $\dep$ of CDs, with $\dep = \idep
\cup \kdep$, where $\idep$ and $\kdep$ are sets of IDs and KDs respectively.
Let $q$ be a CQ over $\R$; we construct $\progeq$, $\progid$ and $\progkd$ in
the following way.

\paragraph{Encoding equalities.}
We introduce a binary predicate $\predeq/2$ that simulates the equality
predicate; to enforce reflexivity, symmetry and transitivity respectively, we
introduce in $\progeq$ the rules
\begin{compactenum}[\itshape (a)]
\item $\predeq(X_i,X_i) \la r(\dd{X}{n})$ for all $r/n$ in $\R$ and for all
  $i\in\{1,\ldots,n\}$
\item $\predeq(Y,X) \la \predeq(X,Y)$
\item $\predeq(X,Z) \la \predeq(X,Y), \predeq(Y,Z)$
\end{compactenum}
Similar rules for encoding equalities are found, for instance, in~\cite{DuLe97,GoNa08}.

\paragraph{Encoding key dependencies.}  
For every KD $\key{r}=\{k\}$ (notice from Section~\ref{sec:conceptual-model}
that in the case of CDs all keys are unary if the original EER schema contains no attributes), with $R$ of arity $n$, we
introduce in $\progkd$ the rule
\begin{eqnarray*}
  \predeq(X_i,Y_i) &\la& r(X_1, \ldots, X_{k-1}, X_k, X_{k+1}, \ldots, X_n),\\
  && r(Y_1, \ldots, Y_{k-1}, Y_k, Y_{k+1}, \ldots, Y_n), \predeq(X_k,Y_k)
\end{eqnarray*}
for all $i$ s.t.{} $1 \leq i \leq n$, $i \neq k$.

\paragraph{Encoding inclusion dependencies.}
The encoding of a set $\idep$ of IDs into a set $\progid$ of rules is done in
two steps.  Similarly to~\cite{CCDL01e,Cali03t}, every ID is encoded by a
logic programming rule $\progid$ with function symbols, appearing in Skolem
terms that replace existentially quantified variables in the head of the
rules; intuitively, they mimic the fresh constants that are added in the
construction of the chase.  We consider the four cases that are possible for an ID
$\sigma$ in a set of CDs coming from an EER schema without attributes:
\begin{compactenum}[\itshape (1)]\label{pag:id-encoding}
	\item $\sigma$ is of the form $r_1[1] \subseteq r_2[1]$, with $r_1/1,\,r_2/1$:
	  we add to $\progid$ the rule $r_2(X) \pmil r_1(X)$.
	\item $\sigma$ is of the form $r_1[k] \subseteq r_2[1]$, with $r_1/n,\,r_2/1,
	  1 \leq k \leq n$: we add to $\progid$ the rule $r_2(X_k) \pmil
	  r_1(\dd{X}{n})$.
	\item $\sigma$ is of the form $r_1[1,\ldots,n] \subseteq r_2[j_1,\ldots,j_n]$,
	  with $r_1/n,\,r_2/n$, and where $(j_1,\ldots,j_n)$ is a permutation of
	  $(1,\ldots,n)$: we add to $\progid$ the rule\\ $r_2(X_{j_1}, \ldots,
	  X_{j_n}) \pmil r_1(X_1, \ldots, X_n)$.
	\item $\sigma$ is of the form $r_1[1] \subseteq r_2[k]$, with $r_1/1,\,r_2/n,
	  1 \leq k \leq n$: we add to $\progid$ the rule $r_2(f_{\sigma,1}(X), \ldots,
	  f_{\sigma,k-1}(X), X, f_{\sigma,k+1}(X), \ldots, f_{\sigma,n}(X)) \pmil
	  r_1(X)$.
\end{compactenum}
Note that in (4) we have used subscripts of the form $\sigma,j$ so as to indicate that for every dependency and for every attribute of $r_2$ there is a different function symbol.

\begin{example}\label{exa:encoding-dependencies}
	Consider the dependencies that do not involve attributes ($\sigma_4$--$\sigma_{13}$) from Example~\ref{exa:ER-query}.
	They can be encoded as follows.
	$$\begin{array}{rrcl}
		\sigma_4:&\rel{employee}(X)&\la& \rel{works\_in}(X,Y)\\
		\sigma_5:&\rel{dept}(Y) &\la& \rel{works\_in}(X,Y)\\
		\sigma_6:&\rel{manager}(X) &\la& \rel{manages}(X,Y)\\
		\sigma_7:&\rel{dept}(Y)&\la& \rel{manages}(X,Y)\\
		\sigma_8:&\rel{employee}(X)&\la& \rel{manager}(X)\\
		\sigma_9:&\rel{works\_in}(X,Y) &\la& \rel{manages}(X,Y)\\
		\sigma_{10}:&\rel{works\_in}(X,f_{\sigma_{10},2}(X))&\la& \rel{employee}(X)\\
		\sigma_{11}:&\rel{manages}(X,f_{\sigma_{11},2}(X))&\la& \rel{manager}(X)\\
		\sigma_{12}:&\predeq(Y_1,Y_2) &\la& \rel{works\_in}(X_1,Y_1), \rel{works\_in}(X_2,Y_2), \predeq(X_1, X_2)\\
		\sigma_{13}:&\predeq(Y_1,Y_2) &\la& \rel{manages}(X_1,Y_1), \rel{manages}(X_2,Y_2), \predeq(X_1, X_2)
		
  \end{array}$$
\end{example}
\paragraph{Query maquillage.}
Since we need to deal with equalities among values in a uniform way, we need
some maquillage (that we call \emph{equality maquillage}) on $q$: replace
every term $t$ in $\body{q}$, with a new variable $X$ not occurring elsewhere
in $q$, and add (as a conjunct) to $\body{q}$ the atom $\predeq(X,t)$.
Henceforth, we shall denote with $\Qeq$ the query after the equality
maquillage.  For example, the query $q(X)\la r(X,c,Y),s(Y)$ becomes $q(X)\la
r(A,B,C),s(D), \predeq(A,X), \predeq(B,c), \predeq(C,Y), \predeq(D,Y)$.

We shall now state that the encoding of CDs by means of the above rules
captures the correct manipulation of facts that is done in the chase (that, we
remind the reader, represents the inference of information done starting from
the initial data and the CDs, under the sound semantics).
In order to do that, in Theorem~\ref{the:answering-prog} below, we first need
to introduce a few auxiliary constructions and lemmata.

We introduce a variant of the chase with equality predicates, denoted
$\chaseeq{\dep}{D}$, that is built as follows from a database $D$ and a set of
CDs $\dep$.
\begin{compactenum}[\itshape (1)]
	\item Add all atoms of the form $\predeq(c,c)$, at level $0$, where $c$ is a constant occurring in $D$.
	\item Include all the facts in $D$ and proceed as for $\chase{\dep}{D}$, but
	\begin{compactenum}[\itshape (a)]
		\item A KD is applicable if there is a key constraint $key(r) = \{k_1,\ldots,k_n\}$ and the chase result constructed so far contains the facts $r(t), r(t')$, and $\predeq(\alpha_1, \beta_1), \ldots, \predeq(\alpha_n, \beta_n)$, with $\alpha_i = t[k_i]$ and $\beta_i= t'[k_i]$.
		When applying the KD rule, instead of merging tuples by replacing the two constants $\alpha_i$ and $\beta_i$ by a combined symbol, add the atoms $\predeq(\alpha_i,\beta_i)$, $\predeq(\beta_i,\alpha_i)$ and all the $\predeq$ atoms that can be derived from the existing ones by transitivity; the level of these $\predeq$ atoms is the same as the lower of the two facts that fired the rule.

	  \item An ID rule is applicable if there is an ID $r[k_1,\ldots,k_n]\subseteq s[j_1,\ldots,j_n]$ such that the chase result constructed so far contains the fact $r(t)$ but there is no fact $s(t')$ such that, for every $i$ such that $1\leq i\leq n$, $\predeq(t[k_i],t'[j_i])$ is in the chase result constructed so far.
		When applying the ID rule, add the atom $\predeq(\alpha,\alpha)$ for each new fresh constant $\alpha$ in the newly introduced fact; the level of $\predeq(\alpha,\alpha)$ is the same as the level of the new fact.

	  \item Whenever an atom of the form $\predeq(c_1, c_2)$ is added, where $c_1, c_2 \in \dom$, and $c_1\neq c_2$, stop the chase procedure (the chase fails).
	\end{compactenum}
\end{compactenum}

\begin{example} \label{exa:chase-eq} Consider again the EER schema of
  Example~\ref{exa:ER-query} and the initial (incomplete) database $D=\{\rel{manager}(m), \rel{works\_in}(m,d)\}$ given in Example~\ref{exa:chase}. Then $\chaseeq{\dep}{D}$ consists of $D$ plus the following facts:
\begin{compactitem}
	\item $\predeq(m,m)$, $\predeq(d,d)$ (constants at level $0$)
	\item $\rel{employee}(m)$, $\rel{manages}(m,\alpha_1)$, $\rel{works\_in}(m,\alpha_1)$, $\rel{dept}(\alpha_1)$, where $\alpha_1$ is a fresh constant (applications of ID chase rules)
	\item $\predeq(\alpha_1,\alpha_1)$ (new fresh constants)
	\item $\predeq(\alpha_1,m)$, $\predeq(m,\alpha_1)$ (application of a KD chase rules)
\end{compactitem}\vspace{-.5cm}
\end{example}

It is straightforwardly seen that $\chase{\dep}{D}$ exists if and only if $\chaseeq{\dep}{D}$ exists.
Clearly, as stated in the following lemma, an isomorphism can be established between the atoms in $\chaseeq{\dep}{D}$ and those in the least Herbrand model of the program consisting of $D$ plus the rules encoding IDs, KDs, and equality.
\begin{lemma}\label{lem:chaseeq-herbrand-one-to-one}
	Consider a database $D$ over a relational schema $\R$ with a set of CDs $\dep=\idep \cup\kdep$, where $\kdep$ and $\idep$ are sets of KDs and IDs respectively, such that $\chase{\dep}{D}$ exists.
	Let $\Pi$ be the program $\progid \cup \progkd \cup \progeq \cup D$ and $M$ its least Herbrand model.
	Then, there is an isomorphism $\mu: \dom \cup \freshdom \ra U_\Pi$, where $U_\Pi$ is the Herbrand universe\footnote{Usually, the Herbrand universe is constructed with respect to a language, but often we can talk about the Herbrand universe of a logic program, intending the Herbrand universe constructed with the constants and function symbols present in that program. The same holds for the notion of Herbrand base.}
of $\Pi$, such that:
\begin{inparaenum}[\itshape (i)]
	\item $\mu(\chaseeq{\dep}{D}) = M$; 
	\item if $\alpha \in \freshdom$ then $\mu(\alpha)$ is a Skolem ground term in $U_\Pi$.
\end{inparaenum}
\end{lemma}
\begin{proof}
	We exhibit the construction of a homomorphism with the desired properties.
	The construction will be inductive on the applications of the immediate consequence operator in the construction of $M$.
	We start from $D$, and we take the identity isomorphism mapping $D$ (as a subset of $\chaseeq{\dep}{D}$) into $D$ (as a subset of $M$).
	Now we consider the following cases of application of the immediate consequence operator, on different kind of rules.
\begin{asparaenum}[\itshape (1)]
	\item \textit{Rule in $\progid$}.
	Assume we are adding a fact $s(\ins{t}_s)$ because of a rule $\rho$ of the form $s(\cdot) \la r(\cdot)$ encoding a dependency $\sigma$ of the form $r[\cdot]\subseteq s[\cdot]$, where $r(\ins{t}_r)$ is a fact in the part $M^*$ of $M$ constructed at a certain point.
	Since, by induction hypothesis, $\mu$ (so far) maps a subset of $\chaseeq{\dep}{D}$ to $M^*$, we take $\mu^{-1}(r(\ins{t}_r))$, which is of the form $r(\ins{u}_r)$: by application of the ID chase rule on $\sigma$ (encoded by $\rho$), we get the addition of a fact $s(\ins{u}_s)$.
	Now extend $\mu$ by adding to it $\{ \ins{u}_s[i] \ra \ins{t}_s[i] \}$
	for every $i$ such that $\ins{u}_s[i]$ is a newly introduced fresh constant (or, equivalently, the corresponding argument in $\rho$'s head contains a Skolem term).
	\item \textit{Rule in $\progkd$}.
	The construction is the same as above, where the added fact in $M^*$ is of the form $\predeq(t_1, t_2)$, with $\{t_1,t_2\} \subseteq U_\Pi$, and  the one in $\chaseeq{\dep}{D}$ is of the form $\predeq(u_1,u_2)$, with $\{u_1,u_2\} \subseteq \dom \cup \freshdom$.
	\item \textit{Rule in $\progeq$}.
	It is straightforwardly seen that rules in $\progeq$ introduce equality atoms, whose corresponding atoms in $\chaseeq{\dep}{D}$ are introduced by enforcing reflexivity, symmetry and transitivity of the predicate $\predeq$, as described in the construction of $\chaseeq{\dep}{D}$.
	The homomorphism $\mu$ is extended accordingly in an obvious way.
\end{asparaenum}
	It is immediate to see that the isomorphism $\mu$ constructed as above is such that values in $\freshdom$ are mapped to Skolem terms (containing function symbols) and vice-versa, and that $\mu(\chaseeq{\dep}{D}) = M$.
\end{proof}
The previous lemma shows an isomorphism between the chase with equalities and
the least Herbrand model of the program comprising the rules for IDs, KDs,
equalities, and the database.  Notice that this result holds for general IDs
and KDs, and not only for CDs: in fact, arbitrary IDs and KDs can be encoded
in the same way we did for CDs.

We then use Lemma~\ref{lem:chaseeq-herbrand-one-to-one} to extend the notion
of level to the atoms of the least Herbrand model: the level of such an atom
is defined as the level of the corresponding (via the isomorphism) atom in the
chase with equalities.

Next, we show that, if we exclude the tuples containing fresh constants, the
answers to a query over the chase coincide with the answers to the query after
maquillage over the chase with equalities.
\begin{lemma}\label{lem:qeq-chaseeq-equiv-q-chase}
  Consider a conjunctive query $q$ over a relational schema $\R$ with a set of
  CDs $\dep=\idep \cup\kdep$, where $\kdep$ and $\idep$ are sets of KDs and
  IDs respectively, and a database $D$ for $\R$, such that $\chase{\dep}{D}$
  exists.  Then the tuples in $\nofresh{\Qeq}(\chaseeq{\dep}{D})$ coincide
  with those in $\nofresh{q}(\chase{\dep}{D})$.
\end{lemma}
\begin{proof}
  By construction of $\chaseeq{\dep}{D}$, if we eliminate all atoms of the
  form $\predeq(\alpha,\beta)$ from $\chaseeq{\dep}{D}$
and replace $\alpha$
  with $\beta$ (or $\beta$ with $\alpha$, provided that the replacing one is
  the fresh constant that lexicographically comes first), we obtain
  $\chase{\dep}{D}$.  We call this process \emph{equality elimination}.
  Suppose that tuple $t$ consisting of non-fresh constants is in $\Qeq(\chaseeq{\dep}{D})$.  Then there exists a
  homomorphism $\mu$ sending $\body{\Qeq}$ to atoms of $\chaseeq{\dep}{D}$ and
  $\head{\Qeq}$ to $t$.  By applying equality elimination to
  $\mu(\body{\Qeq})$ we then obtain atoms in $\chase{\dep}{D}$. These are, in
  turn, an image for a homomorphism $\mu'$ from $\body{q}$ to atoms of
  $\chase{\dep}{D}$.
This can be seen as follows. Consider an atom of the form $\predeq(X,u)$ in $\body{\Qeq}$ such that $\mu(\body{\Qeq})=\predeq(c_1,c_2)$, where $X$ is a variable, $u$ a term, and $c_1,c_2\in\dom\cup\freshdom$. Each time an atom of the form $\predeq(c_1,c_2)$ is eliminated by equality elimination from $\mu(\body{\Qeq})$, remove $\predeq(X,u)$ from $\Qeq$ and replace in it all occurrences of the variable $X$ with the term $u$. At each step of the $\predeq$ elimination process, the two structures are isomorphic; at the end, $\Qeq$ is transformed into a variant of $q$ (i.e., the same as $q$ modulo variable renaming), which proves that $q$ is isomorphic to the result of the equality elimination applied to $\mu(\body{\Qeq})$, i.e., there is the homomorphism $\mu'$ we were looking for.
By construction of $\Qeq$, if $t$ contains no fresh constant, then $\mu'$ necessarily maps $\head{q}$ to $t$.

  For the other inclusion, consider a homomorphism $\mu'$ sending $\body{q}$
  into atoms of $\chase{\dep}{D}$ and $\head{q}$ into $t$.  If the atoms in
  $\mu'(\body{q})$ are in $D$, these are necessarily also in
  $\chaseeq{\dep}{D}$, so all non-$\predeq$ atoms in $\body{\Qeq}$ can also be
  mapped to them by some homomorphism $\mu$; then, the $\predeq$ atoms require the equality of constants in $D$,
  that are necessarily present in $\chaseeq{\dep}{D}$. Then $t$ is also an
  answer in $\Qeq(\chaseeq{\dep}{D})$.
By construction of the $\chaseeq{\dep}{D}$, for every fact $f$ in $\chase{\dep}{D}$ there is a subset $S$ of $\chaseeq{\dep}{D}$, containing only one non-$\predeq$ fact $f'$, such that equality elimination on $S$ yields $f$; we say that $f'$ corresponds to $f$.
 If some atom in $\mu'(\body{q})$ is
  not in $D$, it may have been generated by an ID rule or by a KD rule. 
In the case of an application of an ID rule on a fact $f$ in the chase, then there is a corresponding fact $f'\in\chaseeq{\dep}{D}$ on which the same application is made; note that no tuple merging caused by KD rules in the chase causes new applications of an ID rule.
For a KD rule, in the chase an
  application instantiates fresh constants to other constants from two
  starting tuples; in the chase with equalities, the new tuple is not
  generated, but the two starting tuples remain, and $\predeq$ atoms are
  generated for all merged constants. This means that if an atom
  in $q$ is mapped into such a merged fact, the corresponding (non-$\predeq$) atom in $\Qeq$
  can still be mapped into any of the two starting tuples. By construction of
  $\Qeq$, the body of $\Qeq$ contains one $\predeq$ atom per term in $q$, so
  that each such term can be equalled to the replacing constant in the KD rule
  application (or be left unchanged by mapping the $\predeq$ atom to one that
  equals the term to itself).
\end{proof}
    
With an argument similar to the one used in the proof of Theorem~\ref{the:maximum-level}, it can be shown that, also for the chase with equality, $\maxlevel$ levels are sufficient for query answering. This result is stated below as a corollary of Theorem~\ref{the:maximum-level}.
\begin{corollary}\label{cor:max-level-chase-eq}
  Let $D$ be a database for a relational schema $\R$, $\dep$ a set of CDs over
  $\R$ such that $\chase{\dep}{D}$ exists, and $q$ a conjunctive query over
  $\R$.
  Then, for every tuple $t\in \nofresh{q}(\chaseeq{\dep}{D})$, there exists a homomorphism
  $\mu$ sending $\body{q}$ to facts of $\chaseeq{\dep}{D}$ and $\head{q}$ to $t$
  such that all the atoms in $\mu(\body{q})$ are in the first $\maxlevel$
  levels of $\chaseeq{\dep}{D}$, where $\maxlevel$ is as in Theorem~\ref{the:maximum-level}.
\end{corollary}

Now we can show the main result of this subsection as a consequence of the
previous results.  This result validates our encoding of inclusion
dependencies, key dependencies and equalities into $\progid$, $\progkd$,
$\progeq$ and the query maquillage that returns $\Qeq$ from $q$.  Indeed, if
we put together $\progid$, $\progkd$, $\progeq$ and $\Qeq$ into a program
$\prog_{\Qeq}$, and we evaluate it over a set $D$ of ground atoms, discarding
the answer tuples that contain function symbols, we get exactly the certain
answers to $q$, evaluated over $D$ under $\idep \cup \kdep$.

\begin{theorem} \label{the:answering-prog} Consider a conjunctive query $q$
  over a relational schema $\R$ with a set of CDs $\dep=\idep \cup\kdep$,
  where $\kdep$ and $\idep$ are sets of KDs and IDs respectively, and a
  database $D$ for $\R$, such that $\chase{\dep}{D}$ exists.  Let $\prog$ be
  the set of Horn clauses $\Qeq \cup \progid \cup \progkd \cup \progeq$ and
  let $\progff_{\Qeq}(D)$ be the largest function-free subset of
  $\prog_{\Qeq}(D)$. 
  Then $\progff_{\Qeq}(D) = \ans{q}{\dep}{D}$.
\end{theorem}
\begin{proof}
  By Lemma~\ref{lem:chaseeq-herbrand-one-to-one}, we know that, if we exclude
  the atoms with predicate $\Qeq$, the least Herbrand model $M$ of $\prog \cup
  D$ coincides with $\chaseeq{\dep}{D}$ modulo an isomorphism that sends the
  fresh constants into Skolem terms, and the non-fresh constants into
  themselves.  Therefore, $\prog_{\Qeq}(D)$ coincides with the answers in
  $\Qeq(\chaseeq{\dep}{D})$, modulo this isomorphism; moreover,
  $\progff_{\Qeq}(D)$ coincides with $\nofresh{\Qeq}(\chaseeq{\dep}{D})$,
  since, because of the bijection, atoms with fresh constants correspond to
  atoms with Skolem terms, and vice versa.
	
  By Lemma~\ref{lem:qeq-chaseeq-equiv-q-chase}, we know that
  $\nofresh{\Qeq}(\chaseeq{\dep}{D})=\nofresh{q}(\chase{\dep}{D})$.

  Finally, Theorem~\ref{the:answering-on-chase} guarantees that
  $\nofresh{q}(\chase{\dep}{D}) = \ans{q}{\dep}{D}$, which concludes the
  proof.
\end{proof}

The above result is crucial because it shows the correctness and completeness
of the encoding of the constraints into logic programming rules.

In the next subsection we show how to eliminate the function symbols from
$\prog$, thus obtaining a program expressed in pure Datalog.

\subsection{Elimination of function symbols}
\label{sec:elim-funct}

Now, we want to transform the set of rules $\prog$ of
Theorem~\ref{the:answering-prog} into another set which has pure Datalog rules
without function symbols.  The reason to do so is that in this way we can take
advantage of efficient Datalog engines, while evaluating logic programs with
function symbols would certainly be an overkill.

To do that, we adopt a strategy somehow inspired by the elimination of
function symbols in the \emph{inverse rules} algorithm~\cite{DuGe97} for
answering queries using views.  The problem here is more complicated, due to
the fact that function symbols may be arbitrarily nested in the least Herbrand
model of the program.  The idea here is to rely on the fact that there is a
finite number $\maxlevel$ of levels in the chase that is sufficient to answer
a query, as stated in Theorem~\ref{the:maximum-level}.
We shall construct a Datalog program that mimics \emph{only} the first
$\maxlevel$ levels of the chase, so that the function symbols that it needs to
take into account are nested up to $\maxlevel$ times.  The strategy is based
on the ``simulation'' of facts with function symbols in the least Herbrand
model of $\prog \cup D$ (where $D$ is an initial incomplete database) by means
of ad-hoc predicates that are annotated so as to represent facts with function
symbols.


\begin{definition}[Annotation, annotated predicate, annotated version of an atom]\label{def:annotation}
    Let $\atom{A}$ be an atom of the form $r(\dd{t}{n})$, where every term $t_i$ is of the form $f_{i,1}(f_{i,2}(\ldots f_{i,m_i}(\theta_i) \ldots ))$, every $f_{i,j}$ is a unary function symbol, and every $\theta_i$ is either a constant in $\dom\cup\freshdom$ or a variable.
	The sequence $\bar{\eta}=\dd{\eta}{n}$, with $\eta_i = f_{i,1}(f_{i,2}(\ldots f_{i,m_i}(\bullet) \ldots ))$, is called the \emph{annotation} of $\atom{A}$. 
	The new $n$-ary predicate $r^{\bar{\eta}}$ is called the \emph{annotated predicate} for $\atom{A}$, and
	the function-free atom $r^{\bar{\eta}}(\dd{\theta}{n})$ is called the \emph{annotated version} of $\atom{A}$.
\end{definition}
\begin{example}
	The annotated version of the atom
	$\rel{works\_in}(X,f_{\sigma_{10},2}(X))$
	occurring in the head of rule $\sigma_{10}$ in Example~\ref{exa:encoding-dependencies} is
	$\rel{works\_in}^{\bullet,f_{\sigma_{10},2}(\bullet)}(X,X)$.
\end{example}

Now, to have a program that yields function-free facts as described above,
we construct suitable rules that make use of annotated predicates.  The idea
here is that we want to take control of the nesting of function symbols in the
least Herbrand model of the program, by explicitly using annotated predicates
that represent facts with function symbols; this is possible since we do that
only for the (ground) atoms that mimic facts that are in the first $\maxlevel$
levels of the chase of the incomplete database.  Here we make use of the fact,
proved in Lemma~\ref{lem:chaseeq-herbrand-one-to-one}, that the least Herbrand
model of $\progid \cup \progkd \cup \progeq \cup D$ coincides with
$\chaseeq{\dep}{D}$, modulo renaming of the Skolem terms into fresh constants.
Therefore, we are able to transform a (part of a) chase into the corresponding
(part of the) least Herbrand model.

To do so, we construct a ``dummy chase'', and transform it, in the following
way.
\begin{definition}[Dummy database, dummy chase, dummy chase rules]\label{def:dummy-chase-program}
	Consider a relational schema $\R$ with a set $\idep$ of IDs. 
	\begin{asparaenum}[\itshape (1)]
		\item Let $B$ be a database for $\R$ consisting of exactly one fact of the form $r(\dd{c}{n})$ for every relation $r/n \in \R$, where $\dd{c}{n}$ are distinct constants such that no constant occurs in more than one fact\footnote{It does not matter whether they are fresh or non-fresh, since they will disappear at the end of the process.}; $B$ is called the \emph{dummy database} for $\R$.
		\item Let $\fchase{\idep}{B}$ denote the initial segment of $\chase{\idep}{B}$ consisting of the first $\maxlevel$ levels;  $\fchase{\idep}{B}$ is called the \emph{dummy chase} for $\R$ and $\idep$.
		\item Let $\H$ be as $\fchase{\idep}{B}$, but where each fact (possibly containing fresh constants) is replaced with the corresponding atom (possibly containing function symbols) in the least Herbrand model of $\progid \cup B$; note that such a correspondence exists by Lemma~\ref{lem:chaseeq-herbrand-one-to-one}, because without KDs, if we exclude the $\predeq$ atoms, $\chase{\idep}{B}$ and $\chaseeq{\idep}{B}$ coincide.
		\item Let $\H'$ be as $\H$, but where every atom is replaced with its annotated version.
		\item We denote with $\progdc$ the set of all rules of the form $\atom{A}_2' \la \atom{A}_1'$ such that
		\begin{inparaenum}[\itshape (a)]
			\item there is an arc $(\atom{A}_1, \atom{A}_2)$ in $\H'$, and
			\item by replacing every distinct constant with a distinct variable in $(\atom{A}_1, \atom{A}_2)$, we obtain $(\atom{A}_1', \atom{A}_2')$.
		\end{inparaenum}
		The rules in $\progdc$ are called \emph{dummy chase rules}.
	\end{asparaenum}\vspace{-.5cm}
\end{definition}

\begin{example} \label{exa:dummy-chase} Consider Example~\ref{exa:ER-query};
  in the dummy chase, we introduce, among the others, the fact
  $\rel{employee(c)}$.  This fact generates, according to the ID $\sigma_{10}:\rel{employee[1] \subseteq \rel{works\_in[1]}}$, the fact
  $\rel{works\_in}(c,f_{\sigma_{10},2}(c))$ (after the transformation of the fresh constants into Skolem terms).
Its annotated version is $\rel{works\_in}^{\bullet,f_{\sigma_{10},2}(\bullet)}(c,c)$.
Therefore, $\progdc$ contains, among the others, the rule $\rel{works\_in}^{\bullet,f_{\sigma_{10},2}(\bullet)}(X,X) \la
  \rel{employee}^{\bullet}(X)$.
%
\end{example}

The dummy chase determines all possible nesting sequences of function symbols
that may occur in the first $\maxlevel$ levels of the least Herbrand model of
the program $\progid \cup \progkd \cup \progeq \cup D$: only IDs generate
function symbols, and the dummy chase produces all possible function symbol
sequences that may occur for every relation.

We next show how to
generate a new annotated, function-free program from $\progid \cup \progkd
\cup \progeq$.  Preliminarily, we need some notation: we denote with
$\varsX[h]$ the $h$-th term of a sequence $\varsX$, and with $\bar{\eta}[h]$
the $h$-th element of an annotation $\bar{\eta}$ (which is in turn a
sequence).

\begin{definition}[Function-free rewriting for conceptual dependencies]\label{def:progfin}
	Consider a conjunctive query $q$ over a relational schema $\R$ with a set of CDs $\dep=\idep \cup\kdep$, where $\kdep$ and $\idep$ are sets of KDs and IDs respectively.
	Let $\progbase$ be the set of all rules, called \emph{base annotation rules}, of the form $r^{\bullet,\dots,\bullet}(\dd{X}{n})\la r(\dd{X}{n})$ for every predicate $r\in\R\cup\{\predeq\}$.

    We define $\progfin{q}{\dep}$ as the set of rules $\progdc\cup\progbase$ plus all possible rules of the form
   	$p_0^{\bar{\eta_0}}(\terms_0) \pmil p_1^{\bar{\eta}_1}(\terms_1), \ldots, p_k^{\bar{\eta}_k}(\terms_k)$
		such that:
		\begin{compactenum}
			\item There is a rule $p_0(\terms_0) \pmil p_1(\terms_1), \ldots, p_k(\terms_k)$ in $\progkd \cup \progeq \cup \Qeq$.
			\item Each annotation element $\bar{\eta}_i[j]$ 
			occurs in some rule in $\progdc$.
			\item If $\terms_i[j]=\terms_{i'}[j']$ 
			then $\bar{\eta}_i[j]=\bar{\eta}_{i'}[j']$.
			
		\end{compactenum}\vspace{-.5cm}
\end{definition}

Base annotation rules are just a convenient renaming that allows us to refer to the annotation $^{\bullet,\dots,\bullet}$ to capture also the facts in the database.
Note that $\progid$ is not included in the program since it is already encoded in $\progdc$ in a function-free fashion.

%
%
%

\begin{example} \label{exa:funct-symb-elim} Consider the dependency
	$$\sigma_{13}:\predeq(Y_1,Y_2) \la \rel{manages}(X_1,Y_1), \rel{manages}(X_2,Y_2), \predeq(X_1, X_2)$$
encoding the KD $\key{\rel{manages}}=\{1\}$ from Example~\ref{exa:ER-query}.
Among the annotations occurring in $\progdc$, we have $f_{\sigma_{10},2}(\bullet)$ and $\bullet$ (note
that $\bullet$ necessarily does), as shown in Example~\ref{exa:dummy-chase}.
Then $\progfin{q}{\dep}$ will include, among others, the rules
\[
\hspace{-1cm}\begin{array}{l}
	\predeq^{\bullet,\bullet}(Y_1,Y_2) \la \rel{manages}^{\bullet,\bullet}(X_1,Y_1), 
	                     \rel{manages}^{\bullet,\bullet}(X_2,Y_2),
	                     \predeq^{\bullet,\bullet}(X_1, X_2)\\
	\predeq^{\bullet,\bullet}(Y_1,Y_2) \la \rel{manages}^{f_{\sigma_{10},2}(\bullet),\bullet}(X_1,Y_1), 
	                     \rel{manages}^{\bullet,\bullet}(X_2,Y_2),
	                     \predeq^{f_{\sigma_{10},2}(\bullet),\bullet}(X_1, X_2)\\
	\predeq^{\bullet,\bullet}(Y_1,Y_2) \la \rel{manages}^{\bullet,\bullet}(X_1,Y_1), 
	                     \rel{manages}^{f_{\sigma_{10},2}(\bullet),\bullet}(X_2,Y_2),
	                     \predeq^{\bullet,f_{\sigma_{10},2}(\bullet)}(X_1, X_2)\\
	\predeq^{f_{\sigma_{10},2}(\bullet),\bullet}(Y_1,Y_2) \la \rel{manages}^{\bullet,f_{\sigma_{10},2}(\bullet)}(X_1,Y_1), 
	                     \rel{manages}^{\bullet,\bullet}(X_2,Y_2),
	                     \predeq^{\bullet,\bullet}(X_1, X_2)\\
	\predeq^{\bullet,f_{\sigma_{10},2}(\bullet)}(Y_1,Y_2) \la \rel{manages}^{\bullet,\bullet}(X_1,Y_1), 
	                    \rel{manages}^{\bullet,f_{\sigma_{10},2}(\bullet)}(X_2,Y_2),
	                    \predeq^{\bullet,\bullet}(X_1, X_2)\\
	\predeq^{\bullet,\bullet}(Y_1,Y_2) \la \rel{manages}^{f_{\sigma_{10},2}(\bullet),\bullet}(X_1,Y_1), 
	                     \rel{manages}^{f_{\sigma_{10},2}(\bullet),\bullet}(X_2,Y_2),
	                     \predeq^{f_{\sigma_{10},2}(\bullet),f_{\sigma_{10},2}(\bullet)}(X_1, X_2)\\
\vdots\vspace{-.5cm}
\end{array}
\]
\end{example}

Now we can state our central theorem.
\begin{theorem} \label{the:answering-prog-no-funct} Let $D$ be a database for
  a relational schema $\R$, $\dep$ a set of CDs over $\R$ such that
  $\chase{\dep}{D}$ exists, and $q$ a conjunctive query over $\R$.  Then,
  $\progfin{q}{\dep}_{\Qeq^{\bullet, \ldots, \bullet}}(D) = \ans{q}{\dep}{D}$.
\end{theorem}

\begin{proof}
  The proof is based on the the fact that the least Herbrand model $M$ of
  $\progfin{q}{\dep} \cup D$ is a representation of the first $\maxlevel$
  levels of the least Herbrand model $M_f$ of $\Qeq \cup \progeq \cup \progid
  \cup \progkd \cup D$.
By Lemma~\ref{lem:chaseeq-herbrand-one-to-one}, the first $\maxlevel$ levels of $M_f$ are isomorphic with the first $\maxlevel$ levels of $\chaseeq{\dep}{D}$.
By Corollary~\ref{cor:max-level-chase-eq}, the (non-fresh) answers to $\Qeq$ over the first $\maxlevel$ levels of $\chaseeq{\dep}{D}$ coincide with those found over the whole $\chaseeq{\dep}{D}$.
By Lemma~\ref{lem:qeq-chaseeq-equiv-q-chase}, the (non-fresh) answers to $\Qeq$
 over 
$\chaseeq{\dep}{D}$ coincide with the (non-fresh) answers to $q$ over 
$\chase{\dep}{D}$, which, by Theorem~\ref{the:answering-on-chase}, coincide with $\ans{q}{\dep}{D}$.
Hence, to prove the thesis, we need to show that there is a correspondence between the facts in
$M$ and those in the  first $\maxlevel$ levels of $M_f$.

  We then represent the atoms in $M$ and those in the first
  $\maxlevel$ levels of $M_f$ as two isomorphic structures.  Consider
  therefore the atoms in $M_f$ as being disposed in levels (as in the
  corresponding chase with equalities).  Every two atoms corresponding to an
  ID rule application are connected by an arc.  An $\predeq$ atom has an
  incoming arc for each corresponding atom in the first rule (in $\progkd$ or
  $\progeq$) that produced it via the immediate consequence operator. If we
  exclude $\predeq$ atoms, $M_f$ is a forest whose roots are the atoms in $D$;
  if we include the $\predeq$ atoms, we have a directed acyclic graph, since
  $\predeq$ atoms may have several parents.  We now show that, for each atom
  $\atom{A}$ of the form $p(\theta_1,\dots,\theta_n)$ in the first $\maxlevel$
  levels of $M_f$ there is an atom $\atom{B}$ of the form
  $p^{\eta_1,\dots,\eta_n}(c_1,\dots,c_n)$ in $M$, where each $\eta_i$ is the
  annotation element corresponding to $\theta_i$ and $c_i$ its innermost constant.
  Consider all the ancestors of $\atom{A}$ in $M_f$.
	
  If $p$ is not the $\predeq$ predicate, there is a path $\atom{A}_0, \dots,
  \atom{A}_m = \atom{A}$ in $M_f$, such that $\atom{A}_i$ is at level $i$ and
  $\atom{A}_i$ is $\atom{A}_{i+1}$'s parent.  We prove the claim by induction.
  As base case, we show that there is an atom $\atom{B}_0$ in $M$
  corresponding to $\atom{A}_0$ and an annotation corresponding to
  $\atom{A}_0$'s predicate and terms in $\progfin{q}{\dep}$; but this is obvious, since
  $\atom{A}_0\in D$ and all atoms in $D$ are also in $M$; besides, they also
  exist in
  $M$ with a $^{\bullet, \dots,\bullet}$ annotation, because of the base
  annotation rules in $\progbase$. As inductive step, assume the claim holds
  for all $\atom{A}_j$ with $j \leq i$ and an annotation corresponding to
  $\atom{A}_i$'s predicate and terms is in $\progfin{q}{\dep}$ (let it be $r_i^{\bar{\eta}}$);
  we show that it also holds for $\atom{A}_{i+1}$. There is an ID that
  generates $\atom{A}_{i+1}$ from $\atom{A}_i$. By inductive hypothesis, since
  we are within the first $\maxlevel$ levels, there must be a rule in
  $\progdc$ corresponding to the ID in question, with an atom with predicate
  $r_i^{\bar{\eta}}$ in the body. The application of the immediate consequence
  operator on that rule will produce, by construction, an atom whose predicate
  annotation matches $\atom{A}_{i+1}$'s predicate and terms, and whose
  constants match $\atom{A}_{i+1}$'s innermost constants.

  If $p$ is $\predeq$, the proof is as above, but instead of a single path,
  there may be multiple paths of the form $\atom{A}_0, \dots, \atom{A}_m =
  \atom{A}$; the above argument can be applied to any of them.  The only
  difference is that, instead of ID rules, $\predeq$ atoms are generated
  either by KD rules in $\progkd$ or by the equality rules in $\progeq$. For
  all such rules (and for all the atoms they are applied to) there are the
  corresponding annotated counterparts in $\progfin{q}{\dep}$ that have been
  added by the algorithm for rule annotation.
	
  This proves that, apart from the $\Qeq$ atoms, all the atoms in the first
  $\maxlevel$ levels of $M_f$ have a corresponding annotated atom in $M$. Now,
  the algorithm for rule annotation has added to $\progfin{q}{\dep}$ all
  possible versions of $\Qeq$ in which the head is annotated
  $\Qeq^{\bullet,\dots,\bullet}$ and the positions in which the same variable occurs in the query
  are annotated in the same way, with all possible annotations occurring in
  the first $\maxlevel$ levels of $M_f$. Therefore the $\Qeq$ tuples in $M_f$
  are contained in the $\Qeq^{\bullet,\dots,\bullet}$ tuples in $M$.
	
  For the other inclusion, we simply need to dispose the atoms in $M$
  according to levels, as we did for the atoms in $M_f$.  Starting from the
  atoms of $D$ in $M$ and the $\predeq$ atoms on constants in $D$, by the base
  annotation rules we obtain the same atoms with annotation
  $^{\bullet,\dots,\bullet}$; these annotated atoms are at level $0$ in $M$;
  the non-annotated atoms are never used by any other rule in
  $\progfin{q}{\dep}$ and can be disregarded.  Every other rule in
  $\progfin{q}{\dep}$, when used by the immediate consequence operator,
  generates an atom (in the head) starting from other atoms (in the body);
  when the generated atom is new, we draw an arc from each body atom to the
  head atom, and give it the level $\ell + 1$, where $\ell$ is the maximum
  level of the body atoms.  The resulting structure is again a directed
  acyclic graph, and from this we can proceed as for the other inclusion and
  prove that for each atom in $M$, a corresponding non-annotated atom exists
  in $M_f$, since every rule produced by the algorithm for rule annotation,
  apart from $\progbase$, is a syntactic variant of rules in $\Qeq \cup
  \progeq \cup \progkd$, and the rules in $\progdc$ mimic the rules in
  $\progid$.
\end{proof}

The above theorem suggests our final strategy for computing the answers to a
conjunctive query $q$ expressed over an EER schema, given a database $D$.
\begin{compactenum}[\itshape (1)]
  \item We derive a set $\dep$ of CDs that represent the EER schema.
  \item We check whether $\chase{\dep}{D}$ exists, as described in the proof of
    Lemma~\ref{lem:chase-exists-decidable}, in time polynomial in $|D|$.
  \item Then, we derive a Datalog rewriting that computes all certain answers to
    $q$, according to Theorem~\ref{the:answering-prog-no-funct}.
  \item Finally, we evaluate the Datalog rewriting on $D$.
\end{compactenum}

\subsection{Considerations on complexity}
\label{sec:complexity}
%
We focus here on \emph{data complexity}, i.e., the complexity w.r.t.{} the
size of the data, that is the most relevant, since the size of the data is
usually much larger than that of the schema.
\begin{proposition} \label{the:complexity-answering-by-rewriting}
  The complexity of computing the certain answers to a CQ over an EER schema
  is polynomial in the size of the data if the size $\sizeConnected$ of the largest connected part in the join graph of the instance of the EER schema is bounded.
\end{proposition}

\begin{proof}
  From a CQ $q$ over an EER schema, given a database $D$, we can proceed as
  follows.
\begin{inparaenum}[\itshape (1)]
	\item We check whether the chase exists, which can be done in polynomial time in the size of $D$ by Lemma~\ref{lem:chase-exists-decidable}; if it does not, then query answering is trivial (all $n$-tuples are in the answer to the query $q$, where $n$ is the arity of $q$); 
	\item we construct a Datalog rewriting for $q$, according to what was explained in the previous pages, which does not depend on $D$ but only on $\sizeConnected$, which is assumed to be bounded; 
	\item we evaluate the rewriting on the data.
\end{inparaenum}
  Since the evaluation of a Datalog program is polynomial in data
  complexity~\cite{DEGV01}, the thesis follows.
\end{proof}



\subsection{Extensions of Results}
\label{sec:extensions}

\paragraph{Dealing with inconsistencies.} First of all, as we mentioned in
Section~\ref{sec:chase-answering}, we have always assumed that the initial,
incomplete database satisfies the KDs derived from the EER schema.  This
assumption does not limit the applicability of our results, since violations
of KDs can be treated in different ways.
\begin{inparaenum}[\itshape (1)]
	\item \emph{Data cleaning}
	(see, e.g.,~\cite{HeSt99}): a preliminary cleaning procedure would eliminate
	the KD violations; then, the results from~\cite{Cali06} ensure that no
	violations will occur in the chase, and we can proceed with the techniques
	presented in the paper. 
	\item  \emph{Strictly sound semantics}:
	according to the sound semantics we have adopted, from the logical point of
	view, strictly speaking, a single KD violation in the initial data makes query
	answering trivial (any tuple is in the answer, provided it has the same arity
	of the query); this extreme assumption, not very usable in practice, can be
	encoded in suitable rules, that make use of inequalities, and that can be
	added to our rewritings.  We refer the reader to~\cite{CaLR03b} for the
	details. 
	\item \emph{Loosely-sound semantics}: this assumption is a
	relaxation of the previous one, and is reasonable in practice.
	Inconsistencies are treated in a model-theoretic way, and suitable
	Datalog$^\neg$ rules (that we can add to our programs without any trouble,
	obtaining a correct rewriting under this semantics) encode the reasoning on
	the constraints.  Again, we refer the reader to~\cite{CaLR03b} for further
	details.
\end{inparaenum}

\paragraph{Adding disjointness.} Disjointness between two classes, which is a
natural addition to our EER model, can be easily encoded by \emph{exclusion
  dependencies (EDs)} (see, e.g.~\cite{Lemb04}).  The addition of EDs to CDs
is not problematic, provided that we preliminarily compute the closure,
w.r.t.~the implication, of KDs and EDs, according to the (sound and complete)
implication rules that are found in~\cite{Lemb04}.  After that, we can proceed
as in the absence of EDs.


\section{Discussion}
\label{sec:discussion}

\paragraph{Summary of results.}
In this paper we have employed a conceptual model based on an extension of the ER model, that we called EER (Extended Entity-Relationship), and we have given its semantics in terms of the relational database model with integrity constraints.
We have thus carved out a relevant class of relational constraints, which is a subclass of the well-known key and inclusion dependencies; such a class is important, because in real-world database design the constraints are directly derived from an ER schema.
In fact, the focus of our contribution is on querying incomplete data under an interesting class of relational constraints, rather than on proposing another query language for EER schemata.
Moreover, we argue that our results are independent of the translation from EER to relational.

We have considered conjunctive queries expressed over EER conceptual schemata, and we have tackled the problem of providing the certain answers to queries in such a setting, when the data are incomplete w.r.t.{} the constraints that encode the conceptual schema.
We have characterized a class of relational constraints, namely conceptual dependencies (CDs), that are able to represent EER schemata.
This class is a subclass of KDs and IDs (in the general case the query answering problem is undecidable~\cite{CaLR03}).
In this way, we have reduced the query answering problem under EER constraints into the equivalent problem of query answering under CDs.

We have provided a query rewriting algorithm that transforms a conjunctive query $q$ into a new (recursive Datalog) query that, once evaluated on the incomplete data, returns the certain answers to $q$.

Finally, we have shown how our results can be extended to more general settings, in particular:
\begin{inparaenum}[\itshape (1)]
	\item EER schema with class disjointness; 
	\item the so-called loosely-sound semantics for incomplete data, that overcomes the limitations of the strictly sound one.
\end{inparaenum}

\paragraph{Related work.}  Several works propose query languages for different flavors of EER schemata~\cite{LaTo94,GrLL93,HoEn92,Thalheim:2000}.
Our query language, which does not introduce novel features or characteristics, relies on a standard translation of EER schemata into relational ones.

As pointed out earlier, query answering in our setting is tightly related to containment of queries under constraints, which is a fundamental topic in database theory~\cite{Chan92,CaDL98,JoKl84,KoVa98}.
\cite{CCDL01e}~deals with conceptual schemata in the context of data integration, but the cardinality constraints are more restricted than in our approach, since they do not include functional participation constraints and is-a among relationships.

Other works that deal with dependencies similar to those presented here are~\cite{CDLM*05,CDLL*06}, which deal with a formalism called \emph{DL-Lite} and based on Description Logic; it is easy to establish a correspondence between EER entities and DL-lite \emph{concepts}, and between EER relationships and DL-lite (binary) \emph{roles}.
However, the set of constraints considered in the above works is not comparable to CDs: while it contains some constructs not expressible in EER, on the other hand it is unable to represent, for instance, the is-a among relationships, which we believe is the major source of complexity in the query answering problem.
%
%
Also~\cite{OrCE06} addresses the problem of query containment using a formalism for the schema that is more expressive than the one presented here; the problem is proved to be co\textsc{NP}-hard.
%
In~\cite{CaDL98}, the authors address the problem of query containment for queries on schemata expressed in a formalism that is able to capture our EER model; in this work it is shown that checking containment is decidable and its complexity is exponential in the number of variables and constants of $q_1$ and $q_2$, and doubly exponential in the number of existentially quantified variables that appear in a cycle of the \emph{tuple-graph} of $q_2$ (we refer the reader to the paper for further details).
Since the complexity is studied by encoding the problem in a different logic, it is not possible to analyze in detail the complexity w.r.t.{} $|q_1|$ and $|q_2|$, which by the technique of~\cite{CaDL98} is in general exponential.
If we export the results of~\cite{CaDL98} to our setting, we get an exponential complexity w.r.t.{} the size of the data for the decision problem\footnote{The decision problem of query answering amounts to deciding whether, given a query $q$ and a tuple $t$, $t$ belongs to the answers to $q$.} of answering queries over incomplete databases.
In our work 
we provide a technique that also serves the purpose of computing all answers to a query in the presence of incomplete data.

Our technique for dealing with the non-repairable violations in the chase is the same as in~\cite{CaLR03}.
This is along the lines of consistent query answering~\cite{ArBC99}; a similar approach is found in~\cite{DBLP:journals/iandc/ChomickiM05}.

\paragraph{Future work.}  As future work, we plan to extend the EER model with more constraints which are used in real-world cases, such as covering constraints or more sophisticated cardinality constraints.
We also plan to further investigate the complexity of query answering, providing a thorough study of complexity, including lower complexity bounds.
Also, we are working on an implementation of the query rewriting algorithm, so as to test the efficiency of our technique on large data sets.

\noindent
\paragraph{Acknowledgments.}  
Andrea Cal\`\i{} was supported by the EPSRC project EP/E010865/1 ``Schema
Mappings and Automated Services for Data Integration and Exchange''.  Davide
Martinenghi was supported by the ``Search Computing''
(SeCo) project, funded by the ERC under the 2008 Call for ``IDEAS
Advanced Grants''.


\bibliography{short-string,CM-TPLP2010}

\end{document}